\documentclass[11pt,preprint2]{aastex}
\usepackage{natbib}
\usepackage{amsmath}
\usepackage{epsfig}
\usepackage{graphicx}
\usepackage{amssymb}
\usepackage{color}
\usepackage{epstopdf}
\usepackage{relsize}
\newcommand{\nw}{nW m$^{-2}$ sr$^{-1}$}

\definecolor{orange}{rgb}{1,0.5,0}
\definecolor{purple}{rgb}{0.7,0,0.7}

\shortauthors{Korngut et al. (The CIBER Collaboration)}

\begin{document}

\slugcomment{Draft version \today; for submission to ApJ.}

\title{Measurements of the Zodiacal Light Absolute Intensity
through Fraunhofer Absorption Line Spectroscopy with CIBER}

\author{P.~M.~Korngut\altaffilmark{1,11}, 
  M.~G.~Kim, 
  T.~Arai,
  P.~Bangale\altaffilmark{5},
  J.~Bock\altaffilmark{1,2},
  A.~Cooray\altaffilmark{6},
  Y.T.~Cheng\altaffilmark{1},
  R.~Feder\altaffilmark{1},
  V.~Hristov\altaffilmark{1},
  A.~Lanz\altaffilmark{7}, D.~H.~Lee\altaffilmark{8},
  L.~Levenson\altaffilmark{1},
  T.~Matsumoto\altaffilmark{9},
  S.~Matsuura\altaffilmark{3}, 
  C.~Nguyen\altaffilmark{5}, 
  K.~Sano\altaffilmark{10}, 
  K.~Tsumura\altaffilmark{4},
  and M.~Zemcov\altaffilmark{5}}

\altaffiltext{1}{Department of Physics, California Institute of
  Technology, Pasadena, CA 91125, USA}
\altaffiltext{2}{Jet Propulsion Laboratory (JPL), National Aeronautics
  and Space Administration (NASA), Pasadena, CA 91109, USA}
\altaffiltext{3}{Department of Physics
School of Science and Technology
Kwansei Gakuin University, Hyogo 669-1337, Japan}
\altaffiltext{4}{Department of Natural Science, Faculty of Science and Engineering,
Tokyo City University, Setagaya, Tokyo, 158-8557, Japan}
\altaffiltext{5}{Center for Detectors, School of Physics and Astronomy, Rochester Institute of Technology, Rochester, NY 14623, USA}
\altaffiltext{6}{Center for Cosmology, University of California,
  Irvine, Irvine, CA 92697, USA}
 \altaffiltext{7}{The Observatories of the Carnegie Institution for Science, Pasadena, CA 91101}
\altaffiltext{8}{Korea Astronomy and Space Science Institute (KASI),
  Daejeon 305-348, Korea}
\altaffiltext{9}{Department of Infrared Astrophysics, Institute of
  Space and Astronautical Science (ISAS), Japan Aerospace Exploration
  Agency (JAXA), Sagamihara, Kanagawa 252-5210, Japan}
\altaffiltext{10}{Department of Space Systems Engineering,
School of Engineering,
Kyushu Institute of Technology, Fukuoka 804-8550 JAPAN}
\altaffiltext{11}{Contact author, pkorngut@caltech.edu}

%%%%%%%%%%%%%%%%%%%%%%%%%%%%%%%%%%%%%%%%%%%%%%%
\begin{abstract}
Scattered sunlight from the interplanetary dust (IPD) cloud in our Solar system presents a serious foreground challenge for spectro-photometric measurements of the Extragalactic Background Light (EBL).  In this work, we report on measurements of the absolute intensity of the Zodiacal Light (ZL) using the novel technique of Fraunhofer line spectroscopy on the deepest 8542~\AA~line of the near-infrared CaII absorption triplet.  The measurements are performed with the Narrow Band Spectrometer (NBS) aboard the Cosmic Infrared Background Experiment (CIBER) sounding rocket instrument. We use the NBS data to test the accuracy of two ZL models widely cited in the literature; the Kelsall and Wright models, which have been used in foreground removal analyses that produce high and low EBL results respectively. We find a mean reduced $\chi^{2}$= 3.5 for the Kelsall model and $\chi^{2}$= 2.0 for the Wright model.  The best description of our data is provided by a simple modification to the Kelsall model which includes a free ZL offset parameter. This adjusted model describes the data with a reduced $\chi^{2}$= 1.5 and yields an inferred offset amplitude of $46 \pm 19$~\nw extrapolated to 12500~\AA. These measurements elude to the potential existence of a dust cloud component in the inner Solar system whose intensity does not strongly modulate with the Earth's motion around the Sun.    
\end{abstract}


%%%%%%%%%%%%%%%%%%%%%%%%%%%%%%%%%%%%%%%%%%%%%%%
\section{Introduction}
\label{S:intro}

% reset the footnote counter
\setcounter{footnote}{0}

Measurements of the absolute intensity and spectrum of the Extragalactic Background Light (EBL) at optical and near infrared (NIR) wavelengths capture the redshifted energy released from all nucleosynthesis and gravitational accretion processes throughout cosmic history \citep{Hauser01}. In addition to known galaxy populations, emission sources that contribute to the EBL include the first stellar objects and primordial black holes. If measured with sufficient precision, the EBL spectrum can be used to constrain models of galaxy formation and evolution, connecting energy density to star formation, metal production, and gas consumption as reviewed in \citet{Cooray16}. The EBL also provides an important cosmic consistency test. It allows for a direct comparison of the measured amplitude in the total aggregate signal and the integrated light from all galaxies (IGL) that can be measured directly from deep photometric surveys of individually detected sources. Any discrepancy implies the presence of additional emission from unaccounted components. Diffuse sources could arise during reionization due to recombination radiation, such as a Lyman-alpha background, as well as more exotic sources such as dark matter particle decays and annihilation.

Precise absolute spectro-photometry of the NIR EBL has proven elusive, predominantly due to the bright foreground from sunlight scattered off Interplanetary Dust (IPD) in our own solar system, commonly referred to as the Zodiacal Light (ZL). Apart from making photometry measurements out of the ecliptic plane or past the orbit of Jupiter \citep{ZemcovHorizons,lauer21,Matsuoka2011}, absolute measurements require the accurate subtraction of the ZL foreground. Recent analyses of New Horizons data taken beyond 42 AU from the sun have reported a potential detection of an excess over the IGL in the optical \citep{lauer21}.

Other groups have sought to quantify the absolute level of the EBL by concentrating on indirect measurements such as searching for the imprint left by the EBL on the spectra of bright gamma ray sources \citep{Desai19}.  Indirect high-energy measurements suffer from their own independent set of systematic errors and are therefore very useful as a consistency check.
	
Data from the NIR photometer aboard NASA's Diffuse InfraRed Background Explorer (DIRBE) in the early 1990s were used to generate a geometrical model of the IPD.  The \citet{Kelsall98} model was generated by characterizing the annual modulation of the ZL signal, arising from variations in the integrated dust column density towards a given background field due to the inclination of the dust cloud with respect to the Earth's orbit.
Space-based absolute photometry measurements which rely on this model have yielded EBL estimates of $\sim$60~\nw at 1.25~$\mu$m \citep{Matsumoto15,Cambresy2001}, although \citet{Matsumoto15} allow for a relative calibration difference. A background of this magnitude is difficult to reconcile with the X-ray background and the present-day abundance of metals \citep{madausilk}.

%suggests that the integrated light from galaxies (IGL) can account for only a minor contribution to the total NIR photon budget of the Universe, as the IGL is estimated from deep source counts to be 7~\nw at the same wavelength \citep{IGL2008}.

\citet{Wright2001} produced a ZL model based on DIRBE data under the assumption that the ZL accounts for the entire sky brightness at 25~$\mu$m.  This model, intended to provide a lower limit on the EBL, produced estimates marginally consistent with the IGL \citep{Levenson2007}.  Recent work from the low resolution spectrometer (LRS) aboard CIBER examined the behavior of the EBL from NIR towards the optical under a variety of foreground assumptions \citep{MatsuuraLRS}.  These results suggested a lower limit EBL that was slightly brighter than the IGL with a spectrum markedly redder than ZL.

In this paper, we employ a Fraunhofer line technique to assess the absolute intensity of the ZL foreground. As the ZL is composed solely of scattered Solar emission shortward of $\sim$3~$\mu$m, Fraunhofer absorption lines with well understood and stable equivalent widths can be used to trace the brightness, based on the Solar spectrum.  By accurately measuring the line depth, one can infer the continuum amplitude of the ZL signal alone, independent of a spectrally flat offset.

The Fraunhofer technique was pioneered by \citet{Dube77} to determine the EBL at optical wavelengths.
More recently, \citet{Bernstein2002I,Bernstein2002II,Bernstein2002III},  used this technique by combining space-based photometry from the Hubble Space Telescope with estimates of the ZL intensity from Fraunhofer line measurements made from the ground.  Initial reports of a significant bright EBL detection were later softened as the result of increased scrutiny on systematics resulting from atmospheric and ground reflectance effects \citep{Mattila2003,Bernstein07}.  This highlights the importance of making absolute photometery measurements from space.  Other recent work has applied Fraunhofer spectroscopy to constrain the dynamics in the IPD \citep{wham}.

In this paper, we report a ZL absolute intensity measurement from the custom designed Narrow Band Spectrometer (NBS) with a band targeting the 8542~\AA $ $ CaII Fraunhofer line.  The instrument is one of four aboard the CIBER sounding rocket payload \citep{Zemcov13}, which was flown successfully four times. The detailed design and sensitivity of the NBS is given in \citet{Korngut13}.
This paper reports the detailed analysis of the CIBER NBS science data and the implications for the intensity of the ZL foreground in the NIR.  
%%%%%%%%%%%%%%%%%%%%%%%%%%%%%%%%%%%%%%%%%%%%%%%
\section{The Narrow Band Spectrometer}
\label{S:NBS}

The CIBER NBS telescope is a refractive wide field camera with a 75~mm primary aperture.  In order to obtain the necessary signal-to-noise on the ZL in a short sounding rocket flight, a large etendue was required, yielding an instantaneous field of view (FOV) of $8.5^{\circ} \times 8.5^{\circ}$ sampled by a 256$\times$256 pixel HgCdTe array\footnote{PICNIC detectors manufactured by Teledyne Technologies (http://www.teledyne.com/)}.  A tipped interference filter in front of the camera generates a narrow bandpass whose central wavelength varies as a function of angle of incidence across the field of view.  In this measurement scheme, the spectrum of a uniformly illuminating source such as the ZL can be obtained by extracting the photocurrent level as a function of the radial distance from the boresight, including a shift in angle equivalent to the amplitude of the filter's tip.  

\begin{figure}
\includegraphics[width=0.45\textwidth]{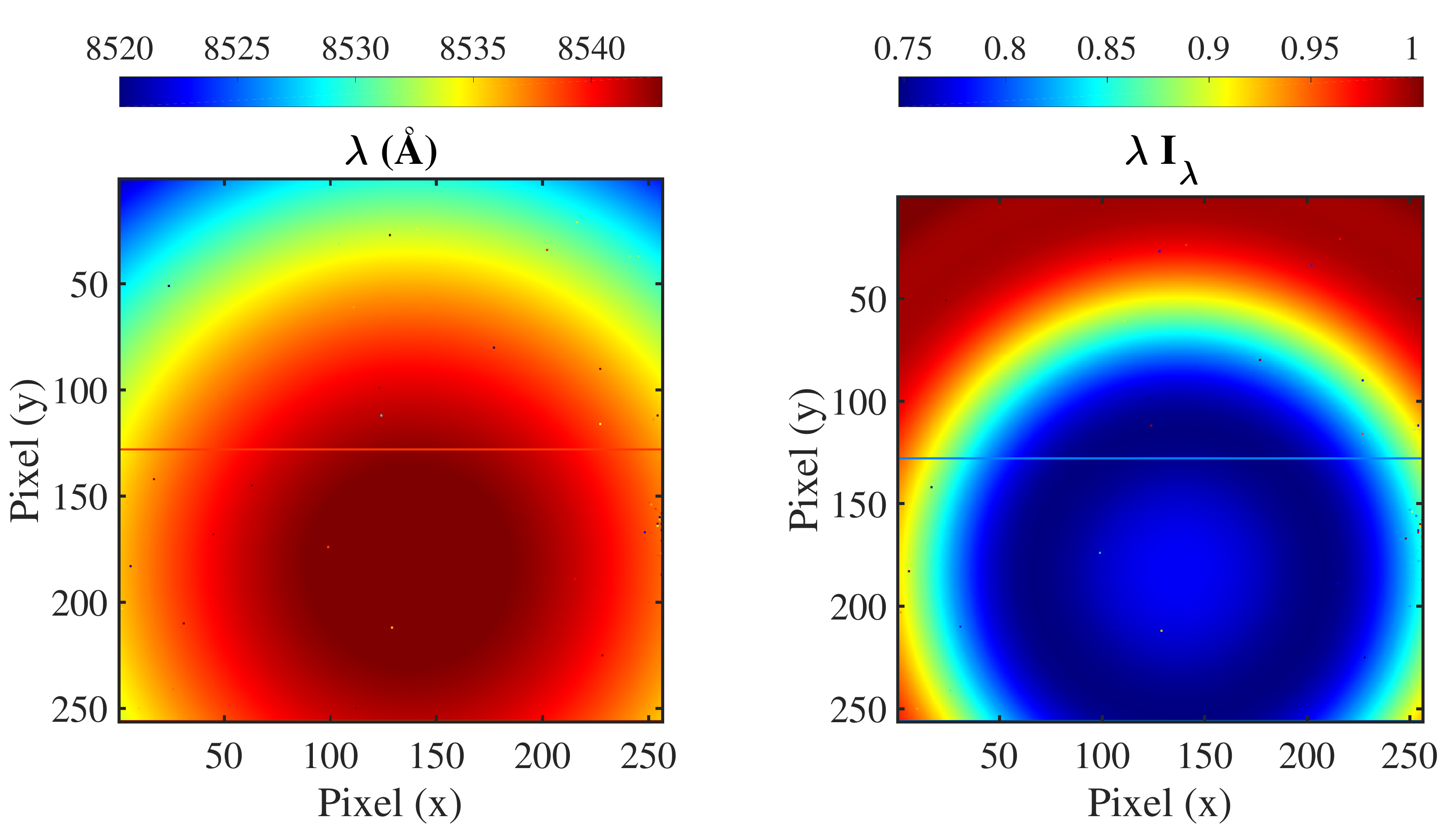}
\caption{Left: The tipped interference filter creates a radially varying central wavelength response across the detector. Right: The continuum normalized Solar absorption spectrum as measured at the focal plane of the NBS.}
\label{fig:nbsill}
\end{figure}

The narrow bandpass was optimized specifically to measure spectral regions both on and off the CaII absorption line without sacrificing sensitivity by extending the spectral range beyond what is necessary to accurately probe the line depth.  The chosen range spans 8520~\AA$< \lambda < $ 8545~\AA $ $ with a resolving power $R = \frac{\lambda}{\Delta\lambda} = 1120$.  In this scheme, an absorption feature in a uniformly illuminating ZL appears as an annular dip.  An illustration of the basic measurement at the NBS focal plane is given in Figure~\ref{fig:nbsill}. For a detailed discussion of the instrument design, see \citet{Korngut13}.

%%%%%%%%%%%%%%%%%%%%%%%%%%%%%%%%%%%%%%%%%%%%%%%
\section{Rocket Flights}
\label{S:rockets}
The data presented here were acquired over three suborbital rocket flights.  The first two were aboard NASA's Black Brant IX two-stage vehicles launched from White Sands Missile Range in New Mexico.  These flights achieved apogees of $\sim$330~km, the first in July, 2010 and the second flown in March, 2012. Both flights displayed nominal performance and the instrument was recovered with no damage. For CIBER's final flight, the payload was flown on a Black Brant XII four stage vehicle, from NASA's Wallops Flight Facility off the coast of Virginia in June, 2013. The Black Brant XII provided a much higher apogee of $\sim$600~km and nearly double the time available for science data acquisition above 250~km, though the payload was intentionally not recovered.  

\section{Field Selection}
\label{S:fields}

We selected fields to optimize a range of criteria, in conjunction with the goals of the other instruments aboard CIBER.  They are located at a range of Galactic and Ecliptic latitudes, modulating the relative contributions of the ZL and other sources of CaII absorption from the Integrated Stellar Light (ISL) and Diffuse Galactic Light (DGL), which are discussed in detail in Sections  \ref{sS:dgl}, ~\ref{sS:BISL}, and ~\ref{sS:FISL}.  The three sub-orbital rocket flights took place during different phases of the Earth's orbit around the Sun.  Due to the inclination of the IPD cloud in the ecliptic plane, fields which have identical extrasolar backgrounds will have different ZL intensities, as the column density we look through varies.  The fields are listed in Table~\ref{table:fields}, along with the predicted 1.25~$\mu$m intensities from the Kelsall and Wright models. 
%%%%%%%%%%%%%%%%%%%%%%%%%%%%%%%%%%%%%%%%%%
\section{Data Reduction}
\label{S:data}
The raw data arriving via radio frequency telemetry consist of a series of timestamped PICNIC detector frames sampled at 4~Hz.  These frames are synchronized with housekeeping and event flags such as cold shutter status, coarse and fine target acquisition indicators and a rocket door status.  Photocurrent images of each field are generated using the sample-up-the-ramp method, or  fitting a linear slope to the integrated charge for each pixel as a function of time.  The first 10 frames after a reset are discarded from analysis to avoid transient response from charge injection.  Before spectral extraction, the data are subject to a suite of processing steps, described in this section.  Detailed information on the ancillary laboratory measurements used to generate various instrument data products can be found in \citet{Korngut13}.

%%%%%%%%%
\onecolumn
\begin{table}[t!]
\centering
\begin{tabular}{c|c|c|c|c|c|c}
\hline
\hline
Field & $\alpha$ & $\delta$  & Flight & Exposure & Kelsall 1.25$\mu$m & Wright 1.25 $\mu$m \\
	& (h) & (deg) &	  & (s) &  (nW~m$^{-2}$~sr$^{-1}$) &  (nW~m$^{-2}$~sr$^{-1}$)\\
\hline
\hline
NEP    	       & 18.06 & 66.10 & July, 2010 & 63  & 233 & 255\\
BootesA           & 14.55 & 34.58 & July, 2010 & 59 & 316 & 349\\
BootesB           & 14.46 & 33.08 & July, 2010 & 32 & 325 & 358\\
Elat30A           & 15.77 & 09.29 & March, 2012 & 24 & 402 & 444\\
BootesB           & 14.45 & 33.02 & March, 2012 & 47 & 327 &  365\\
Elat10           & 12.69 & 08.32 & June, 2013 & 46 & 551 & 605\\
Elat30B           & 12.87 & 28.29 & June, 2013 & 47 & 397 & 447\\
BootesB           & 14.48 & 33.50 & June, 2013 & 52 & 297 & 337 \\
Elais-N1           & 16.19 & 54.34 & June, 2013 & 47 & 242 & 270 \\
\hline
\end{tabular}
\caption{Selected target fields and their ZL intensities from the models.  Data were collected across three sounding rocket flights.  The fields were selected to span a range of both Ecliptic and Galactic lattitudes to modulate the relative contributions of ZL, ISL and DGL. \label{table:fields}}
\end{table}
\begin{table}
\centering
\begin{tabular}{c|c|c|c|c}
\hline
\hline
{\bf Parameter} & {\bf Description} & {\bf Component} & {\bf Ancillary Model} & {\bf Float?} \\
\hline
\hline
$A_{ZL}$ & Amplitude parameter & ZL & - & Yes \\
\hline
$G_{ZL,x,y}$ & Spatial Gradient & ZL & \citet{Kelsall98} & No \\
\hline
$F_{\lambda,ZL}$ & Spectral Template & ZL & \citet{Korngut13} & No \\
\hline
$A_{DGL}$ & Continuum Amplitude & DGL & \citet{araidgl} & No \\
\hline
$G_{DGL,x,y}$ & Spatial Gradient & DGL & \citet{schlegeldust} & No \\
\hline
$F_{\lambda,DGL}$ & Spectrum & DGL & \citet{Lehtinen13} & No \\
\hline
$A_{BISL}$ & Continuum Amplitude & BISL & \citet{dss} & No \\
\hline
$G_{BISL,x,y}$ & Stellar positions & BISL & \citet{dss} & No \\
\hline
$F_{\lambda,BISL}$ & Stellar CaII & BISL & \citet{Lehtinen13} & No \\
 &  &  & \citet{trilegal} &  \\
 \hline
$A_{FISL}$ & Continuum Amplitude & FISL & \citet{trilegal} & No \\
 \hline
 $F_{\lambda,FISL}$ & Faint unresolved stars & FISL & \citet{Lehtinen13} & No \\
      &  &  & \citet{trilegal} &  \\
 \hline
$C$ & Featureless offset & DC & -	& Yes \\
    &  &  AGL & 	& \\
    &  &  EBL & 	& \\

\hline
\end{tabular}
\caption{Model parameter definitions and background used in the analysis.\label{table:params}
}
\end{table}
\twocolumn
%%%%%%%%%%%%%%%%%%%%%%%%%%%%%%%%%%%%%%%%%%%%%%%%%%%
%%%%%%%%%%%%%%%%%%%%%%%%%%%%%%%%%%%%%%%%%%
%%%%%%%%%%%%%%%%%%%%%%%%%%%%%%%%%%%%%%%%%%%%%%%
\subsection{Dark Current}
\begin{figure}
\includegraphics[width=0.45\textwidth]{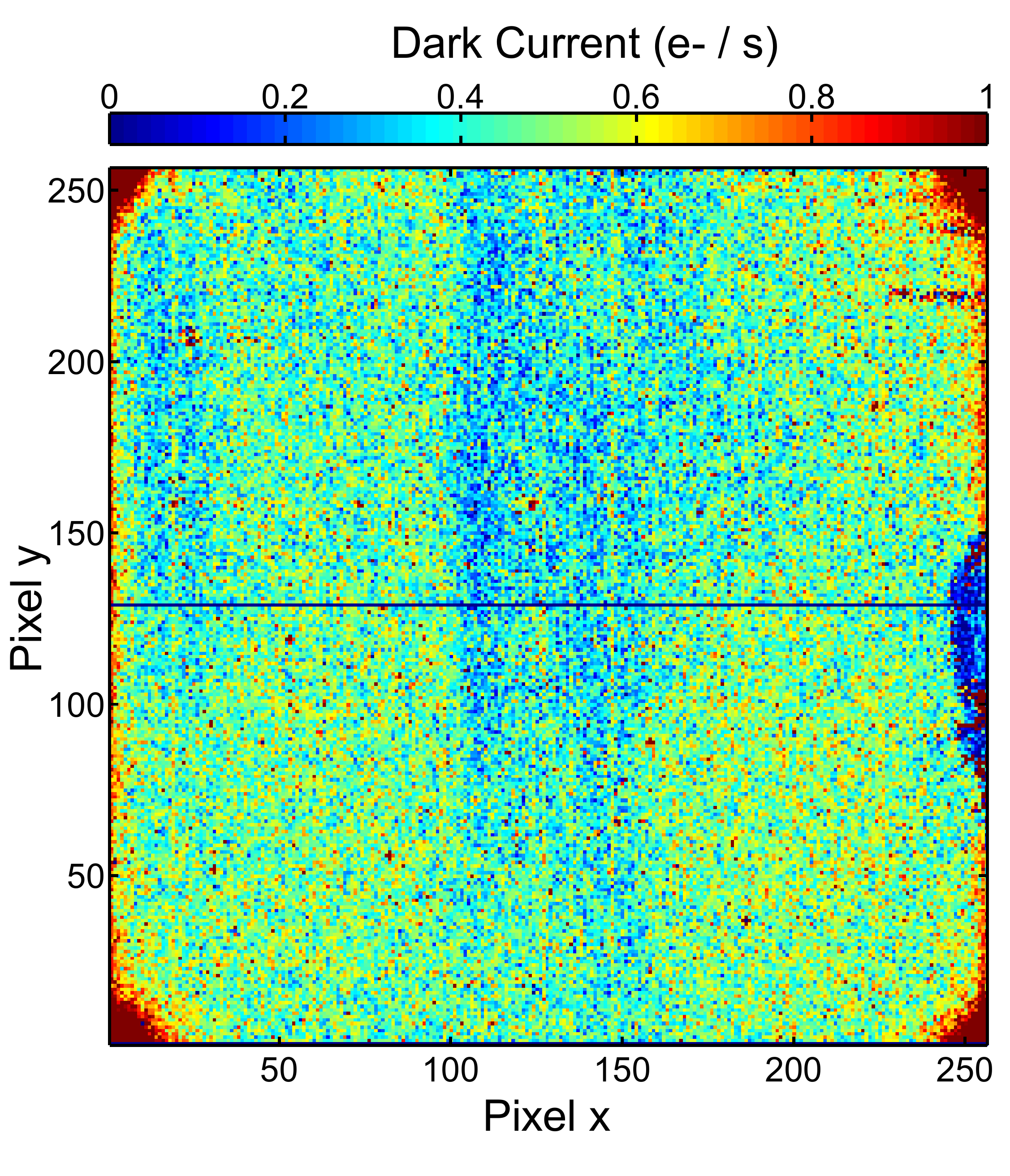}
\caption{NBS dark current template image in array coordinates.}
\label{fig:dc}
\end{figure}

The NBS is equipped with an optical shutter, located just above the detector at 79~K to measure the detector dark current (DC) in-situ. Immediately preceding each flight, a suite of dark images are measured while the rocket is on the rail awaiting launch.  These data are coadded to generate a pixel-wise template of the variations in DC containing negligible read noise.  In addition, during each flight, the shutter is closed for approximately 50~s to obtain an in-flight dark measurement, albeit containing significant read noise contributions.    The average dark current is typically 0.5~e-/s in unmasked pixels, comparable to the photocurrent induced from the ZL. In this measurement, the most important DC feature is the reproducible large scale structure across the array due to fabrication inhomogeneities in the detectors.  The random alignment of array-scale dark current structure and the wavelength distribution produces a bias in fitting the CaII line depth if left unsubtracted.  The rail DC template is therefore subtracted in the map domain before spectral analysis. An image of a typical dark current template is shown in array coordinates in Figure~\ref{fig:dc}.  Bright features at the corners are caused by multiplexer glow and are masked, along with the ovular shaped feature towards the $x = 256, y = 128$ location.

%%%%%%%%%%%%%%%%%%%%%%%%%%%%%%%%%%%%%%%%%%%%%%%
\subsection{Step Removal}
The detector readout system uses two separate boards to read individual halves of the array.  Drifts in the independent amplifier chains of the two channels can introduce different offsets on either side of the array. Since the NBS measurement quantifies the depth of an absorption line, it is not sensitive to arbitrary DC offsets. However, since the radial dispersion of the CaII feature is not perfectly aligned with the mid plane of the detector array, averaging across the entire FOV would introduce spurious inferred depth of the absorption feature.  Therefore, we extract a spectrum independently on each half of the array, calculate the offset between the two spectra and remove the difference.
For each field, spectra for each half of the array are extracted independently.

%%%%%%%%%%%%%%%%%%%%%%%%%%%%%%%%%%%%%%%%%%%%%%%
\subsection{Flat-Field}
\begin{figure}
\includegraphics[width=0.45\textwidth]{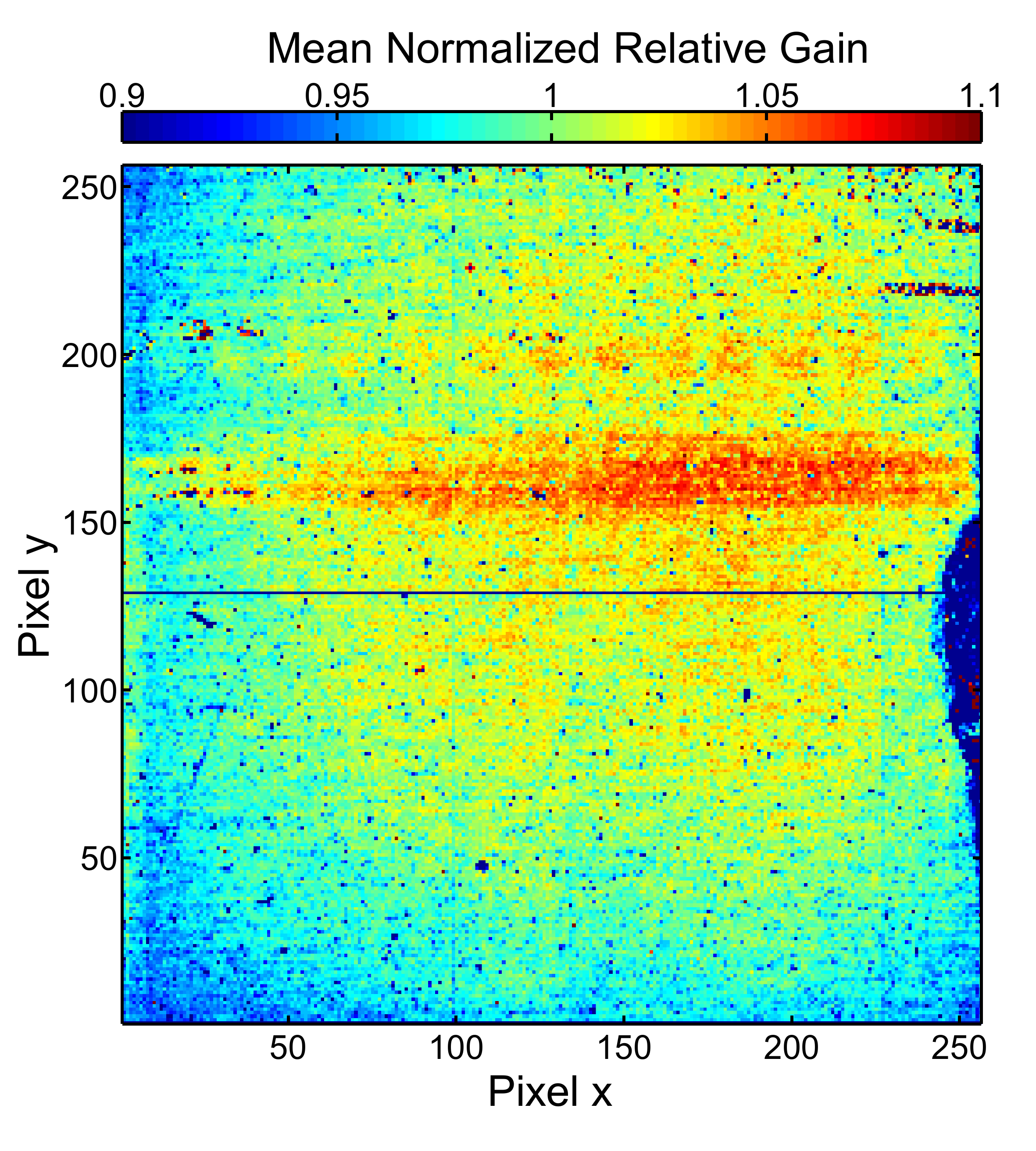}
\caption{Flat-field gain matrix in array coordinates.  The colorscale is normalized by the mean across the array.  No Masking is included in the figure.}
\label{fig:flat}
\end{figure}

The flat-field response of the NBS was measured before and after each flight campaign in the laboratory. The measurement technique is discussed in detail in \citet{Zemcov13} and entails coupling the NBS aperture into an integrating sphere enclosed in a vacuum chamber.  Broadband light is coupled to the integrating sphere via fiber optic cable and the light intensity can be varied as desired. The flat field is constructed by coadding dozens of dark-subtracted exposures together and normalizing the response to the mean of all pixels used in spectral analysis.  The measurement is repeated at three different light levels spanning an order of magnitude in photocurrent.   An example of a flat-field matrix is shown in Figure~\ref{fig:flat}.  The structure in the flat-field matrix comes from a combination of intrinsic quantum efficiency variation in the detector array along with reflections and varied illumination in the optical chain.  Because the optics were disassembled and re-assembled between flights, we use a unique flat field template for each flight.

%%%%%%%%%%%%%%%%%%%%%%%%%%%%%%%%%%%%%%%%%%%%%%%
\subsection{Pixel Masking}
Errant pixels are flagged and excluded from the analysis based on a range of criteria as follows.
%}
\begin{enumerate}

\item Hot and dead pixels are rejected.  These are identified as extreme statistical outliers in dark current measurements.  The specific PICNIC chip in the NBS has a thin arc-like defect $\sim$10 pixels in radius towards the edge of the array which is masked here.  In total, $\sim$3\% of pixels are removed at this stage. This mask is used in common for all fields in a given flight.

\item For each field, a variance estimator map is generated from the statistics of the best line fit in determining the estimated photocurrent.  Statistical outliers which deviate from the mean in excess of 5$\sigma$ are rejected. These account for typically $\sim$1.5\% of the array, including an overlap in population with the previous condition.

\item The corners of the array are masked to avoid contamination from a spurious signal originating from self emission from the detector's multiplexor at the corner of each quadrant.  The regions along the interface of the four quadrants of the array are masked as well.  The effect being mitigated is visible in the dark current image in Figure~\ref{fig:dc}.

\item Pixels with extreme values in the flat-field matrix are excluded (greater than 5$\sigma$ from the mean).
\end{enumerate}

%%%%%%%%%%%%%%%%%%%%%%%%%%%%%%%%%%%%%%%%%%%%%%%
\subsection{Calibration}
\begin{figure}
%\begin{center}
\includegraphics[width=0.45\textwidth]{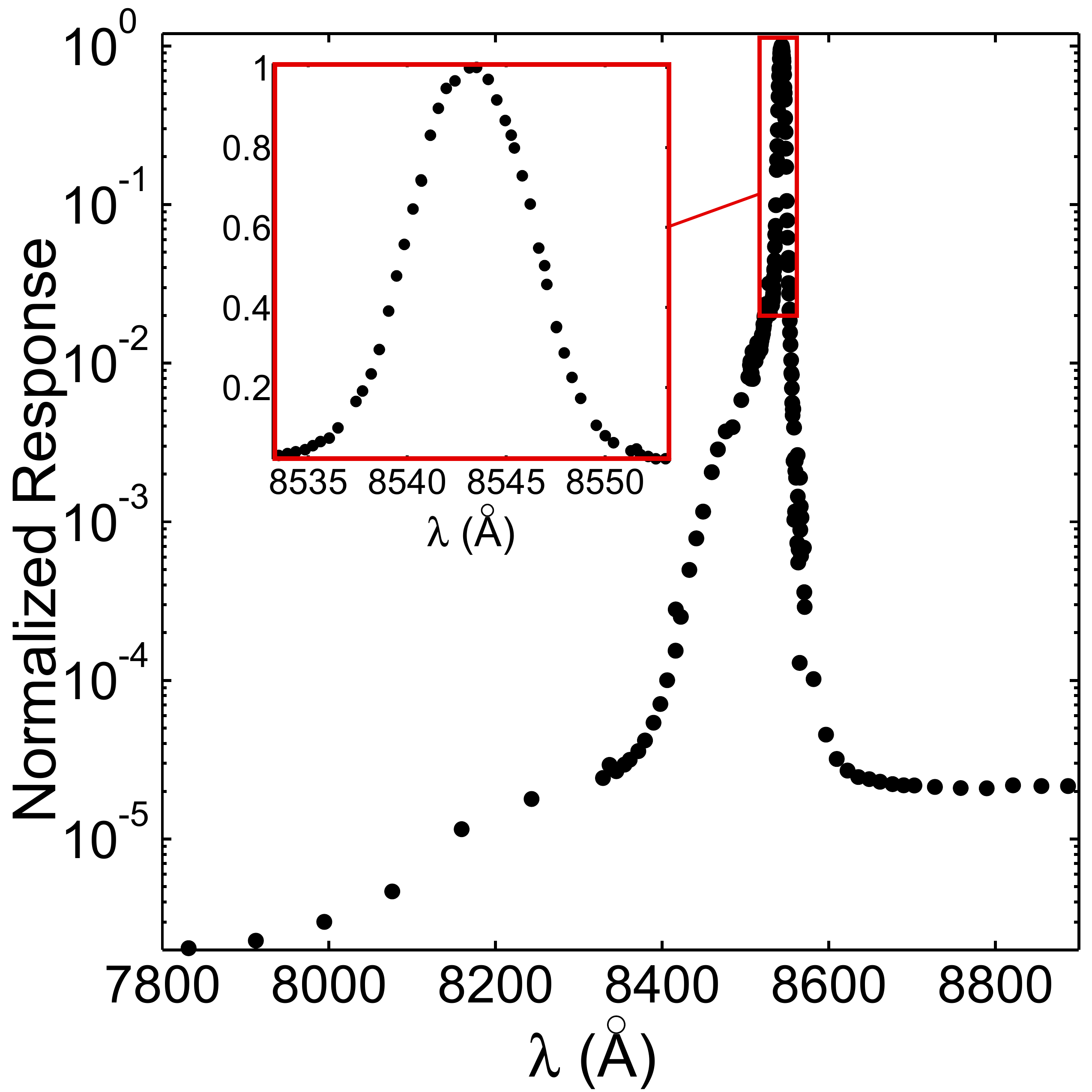}
%\end{center}
\caption{NBS spectral response function measured with the SIRCUS laser facility for a single pixel. The main band is shown on a linear scale in the inset. The out-of-band response is shown in the full plot with a factor of million in dynamic range. This function corresponds to $\Lambda_{163,168}(\lambda)$ in Equation~\ref{eqn:2dmodel}.}
\label{fig:filter}
\end{figure}

The absolute spectro-photometric calibration of the NBS is obtained through a suite of laboratory measurements in collaboration with the National Institute of Standards and Technology (NIST). In particular, their SIRCUS laser facility \citep{Brown06} provides an intensity stabilized monochromatic source with negligible intrinsic linewidth for our purposes. The central wavelength of the laser is tunable and can be scanned across the NBS band. This source is coupled to the NBS aperture through an integrating sphere, with a series of absolutely calibrated radiometers (to 0.2\% accuracy) and monitor detectors in the optical chain.  A measurement of the spectral response function of each pixel is obtained with a dynamic range of $10^6$ by dramatically increasing the intensity of the light in the sphere when probing wavelengths adjacent to the band.  Figure~\ref{fig:filter} shows the measurement on a single pixel over a range of 1000~\AA. 

The spectral response was measured on five separate occasions, spanning numerous intermittent thermal cycles, rocket flights and mechanical adjustments to the payload.  A weighted average across all measurements provides a global calibration factor of CF~=~$631 \pm 15 $~nW m$^{-2} $sr$^{-1} / $ e$^{-}$s$^{-1}$, i.e. reproducible to 2.4\%.  

\subsection{Astrometry}
Astrometric registration is carried out in a two step process.  First, an initial pointing solution is determined from the attitude control system, which specifies a mapping for each flight from pitch, yaw and roll; to right ascension, declination and parallactic angle.  An offset in all three parameters is then fit for using cross-identified stars in the 2MASS all sky catalog \citep{twomass}.  Radial distortion across the FOV is accounted for using detailed optical ray tracing simulations of the instrument, and are validated against star positions across the FOV.

%%%%%%%%%%%%%%%%%%%%%%%%%%%%%%%%%%%%%%%%%%%%%%%
\section{2D Component Modeling}
\label{S:2D}
For tipped filter spectroscopy, the raw measurement consists of an image of the sky in which each pixel has a slightly different bandpass (as depicted in Figure~\ref{fig:nbsill}).  In the limit that the ZL illumination is dominant and uniform across the FOV, the image produced by the NBS would appear as an azimuthally symmetric torus centered around the peak wavelength.  However, in practice there are other components to the total sky signal which have spatial structure that produce a non-zero spurious ZL signal through the fitting process.  In fact, in the NBS's $8.5^{\circ} \times 8.5^{\circ}$ FOV, the ZL itself can have non-negligible gradients, particularly at lower ecliptic latitudes.  

To make the most accurate determination of the ZL intensity from the NBS data set, we model each field in two dimensions
relying on ancillary data for the other components.
The model we generate for the measured brightness of each pixel {\it x,y} in NBS detector array coordinates can be expressed as a sum of integrals given by:

\begin{multline}
\lambda I_{\lambda,total,x,y} = \\
A_{ZL}G_{ZL,x,y} \int d\lambda \Lambda_{x,y} (\lambda) F_{\lambda,ZL}(\lambda)\\
+\\
A_{DGL}G_{DGL,x,y} \int d\lambda \Lambda_{x,y} (\lambda) F_{\lambda,DGL}(\lambda)\\
+\\
A_{BISL}G_{BISL,x,y} \int d\lambda \Lambda_{x,y} (\lambda) F_{\lambda,BISL}(\lambda)\\
+\\
A_{FISL}\int d\lambda \Lambda_{x,y} (\lambda) F_{\lambda,FISL}(\lambda)\\
+\\
C,\\
\label{eqn:2dmodel}
\end{multline}
where $\Lambda_{x,y}$ is the spectral response function of pixel $(x,y)$, $G_{i,x,y}$ encompasses the spatial variation of component $i$ across the FOV, $F_{\lambda,i}$ is the narrow band spectrum of each component, $A_{i}$ is an amplitude normalization parameter and $C$ is a spectrally flat offset which can include emission from residual airglow (AGL) and the EBL which are both spectrally smooth in this region. The offset C can also include electrical effects within the detector and readout system which is why it is not shown in a bandwidth integral.

\begin{figure}
%\begin{center}
\includegraphics[width=0.45\textwidth]{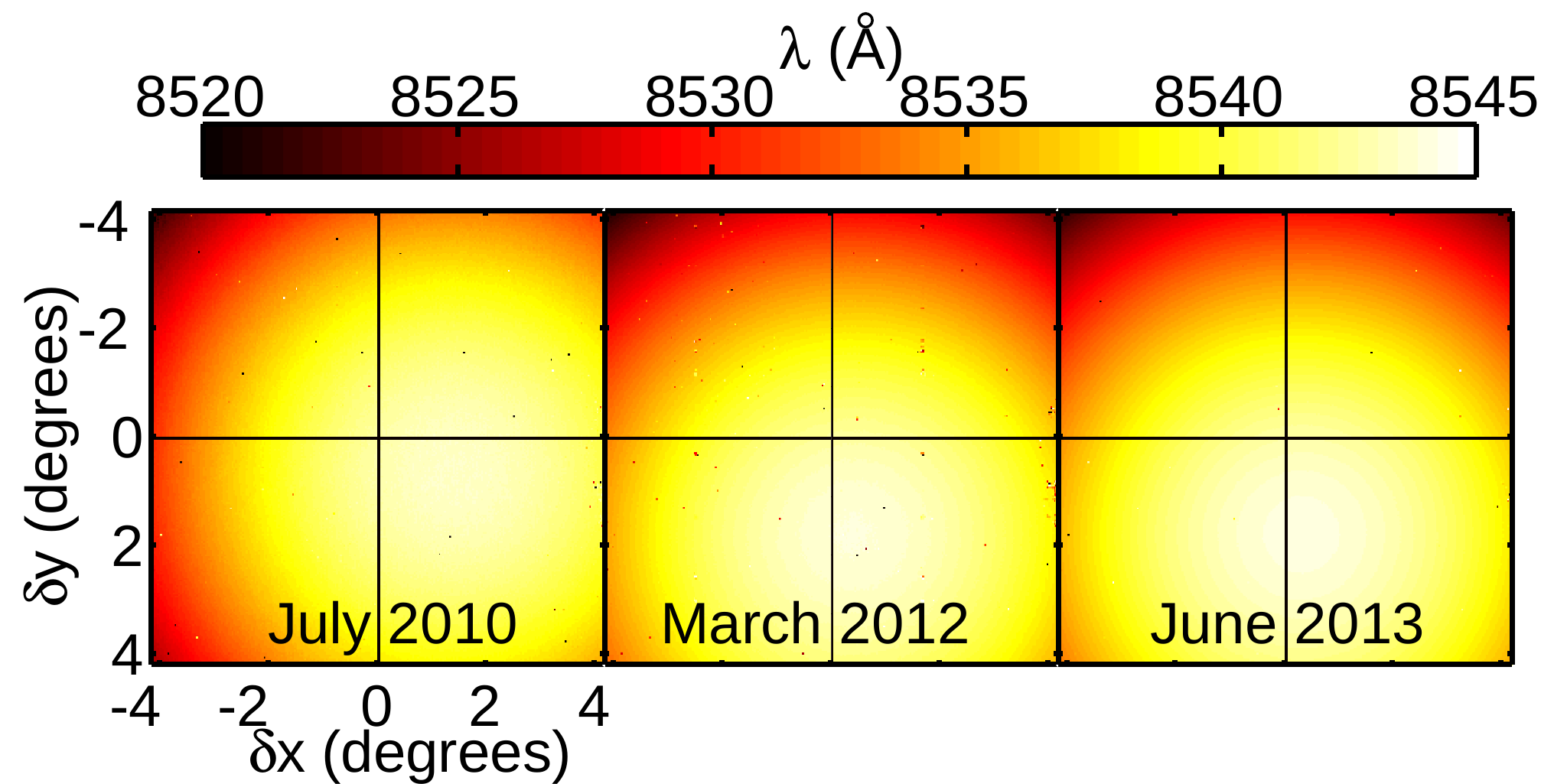}
%\end{center}
\caption{Wavelength calibration maps for the detector array used in analysis of each of the three flights.  The values displayed in each pixel here correspond to the central wavelengths of $\Lambda_{x,y}$ in Equation~\ref{eqn:2dmodel}}
\label{fig:wavemaps}
\end{figure}

 To quantify the impact they have on the measurement, the characteristics of each component must be understood thoroughly. Subsections \ref{sS:isl} through \ref{sS:zl} detail the assumptions in their modeling and Section~\ref{S:sys} describes the propagation of their uncertainties. A summary of the parameters appearing in Equation~\ref{eqn:2dmodel} can be found in Table~\ref{table:params}.  Due to changes in the instrument configuration between flights, consisting mainly of a rotation of the relative alignment of the tip in the band defining filter and the detector array, the values of  $\Lambda_{x,y}$ are unique for each pixel in each flight.  The central wavelengths, as mapped to the array for all cases considered are given in Figure~\ref{fig:wavemaps}.
As discussed in Section~\ref{S:fits}, the only parameters considered free in Equation~\ref{eqn:2dmodel} are $A_{ZL}$ and $C$, all other parameters are constrained by external data or models.

%%%%%%%%%%%%%%%%%%%%%%%%%%%%%%%%%%%%%%%%%%%%%%%

%%%%%%%%%%%%%%%%%%%%%%%%%%%%%%%%%%%%%%%%%%%%%%%
\subsection{Integrated Stellar Light (ISL)}
\label{sS:isl}

Galactic stars contribute significant flux to the measured CaII signal. With coarse $2'$ pixels, the NBS has limited power to detect and mask stars individually, as each pixel contains many sources.  Unlike the EBL, which is sourced from all redshifts, the spectra of the local stellar population adds coherently and therefore contributes to the inferred depth of the CaII line. Ideally, we would rely on deep ancillary star catalogs in each field to mask the images aggressively and remove stellar contamination directly.  However, with the limited spatial resolution of 2 arcminutes, masking NBS data to the necessary depth would result in an intolerable loss of pixels.  We therefore mask only to moderate depth, defined by magnitude $M_{cut}$, and rely on modeling and ancillary data to account for the remaining surface brightness below that threshold.  For stars brighter than $M_{cut}$ (BISL), we rely on an ancillary all-sky catalog at $\lambda = 880$~nm to generate pixel masks.  For the aggregate faint stellar population (FISL), we rely on models of the Galaxy.  The implementation of each is described in detail in the following subsections.

\subsubsection{Bright Stars}
\label{sS:BISL}

 For this study, we use the catalog produced by the USNO-B2 Digital Sky Survey (DSS) \citep{dss}, which includes a NIR band centered at $\lambda = 880$~nm.  While the wavelength difference and resolution between these instruments is less than ideal, it is the closest all-sky catalog publicly available, and we account for the minor wavelength difference through simulations.

Detailed knowledge of the effective PSF is essential for an accurate accounting of the stellar foreground.  The average PSF is measured in each flight independently by stacking on DSS star positions with $7 < M_{AB} < 9$.  As the NBS design under-samples the PSF significantly, the stack is done on a sub-pixel grid, applying the technique implemented in \citet{symons21}. 

The star masking algorithm we apply is characterized by two parameters, an AB cutoff magnitude $M_{cut}$ and a flux threshold parameter $t$. Using the measured PSF along with fluxes and positions from the DSS catalog, model stellar maps for sources brighter than $M_{cut}$ in each field are generated.  The maps are initially generated on a pixel scale four times finer than the NBS native resolution ($30''$ pixels) to account for the sub-pixel centroiding of sources. They are then interpolated onto the $2'$ grid. Pixels in the model maps with values brighter than $t$ are masked in spectral extraction.  This technique naturally removes a larger region around brighter stars. 

\begin{figure}
%\begin{center}
\includegraphics[width=0.5\textwidth]{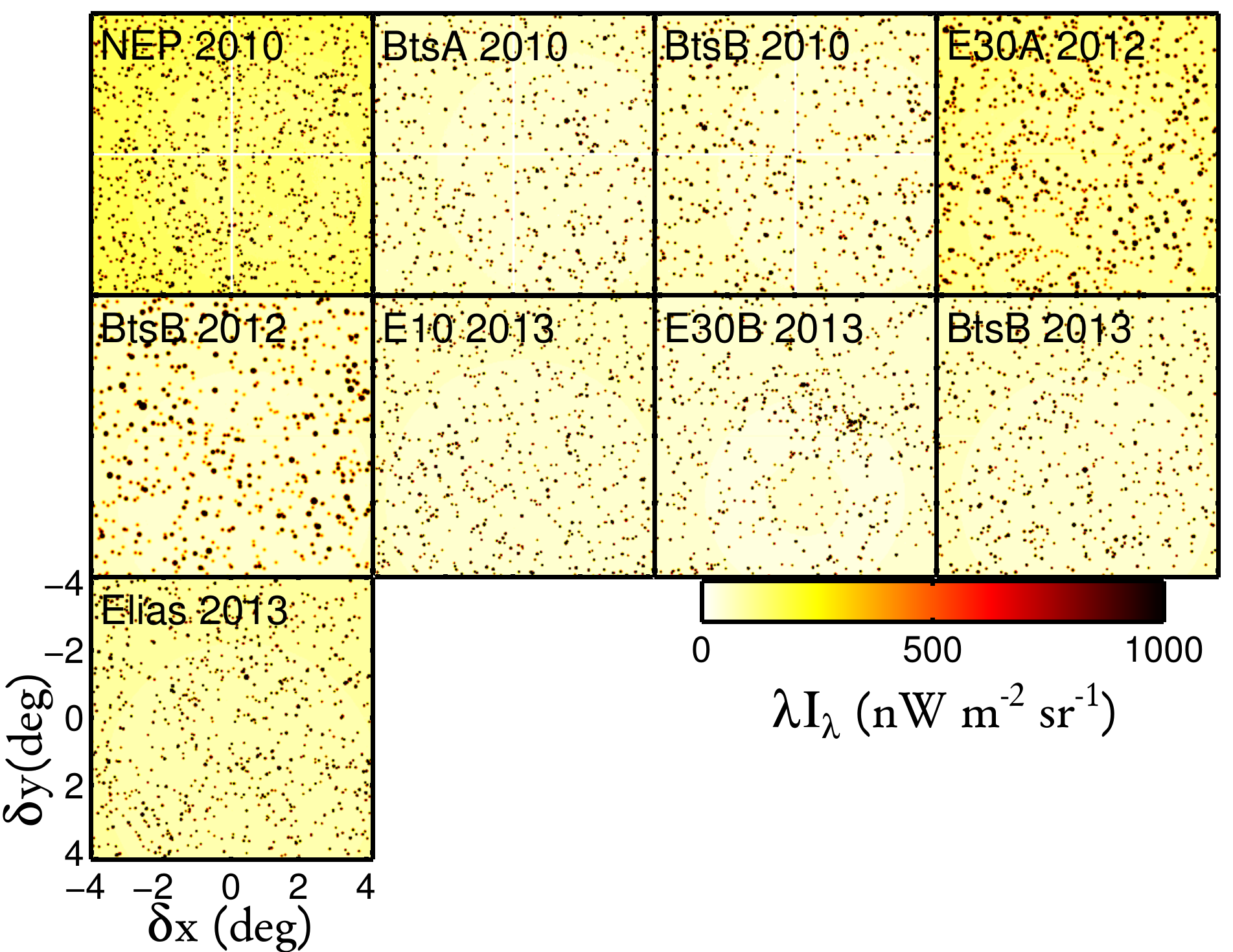}
%\end{center}
\caption{Template images of the integrated stellar light corresponding to the sum of $G_{BISL,x,y} \int d\lambda \Lambda_{x,y} (\lambda) F_{\lambda,BISL}(\lambda)
+ \int d\lambda \Lambda_{x,y} (\lambda) F_{\lambda,FISL}(\lambda)$ in Equation~\ref{eqn:2dmodel}. Individual sources are generated from the DSS catalog positions and fluxes, determined NBS astrometry and a model of the NBS PSF.  The faint background is generated from the Trilegal model, and shown as observed by the NBS wavelength response assuming Solar CaII absorption.}
\label{fig:isl}
\end{figure}

\subsubsection{Faint stars}
\label{sS:FISL}

To account for the integrated emission from stars below the mask threshold, we rely on the Galactic stellar population code TRILEGAL \citep{trilegal}.  This code is run for the CIBER target fields to generate statistically accurate simulated stellar catalogs down to $M_{AB} = 26$.  The simulation calculates both the flux observed using the CIBER NBS filter as well as the DSS-i2 filter.  Simulated noiseless observations of stellar fields are produced for both filters by randomly populating the stars across the NBS FOV using the appropriate PSF model.  

\begin{figure}
\begin{center}
\includegraphics[width=0.4\textwidth]{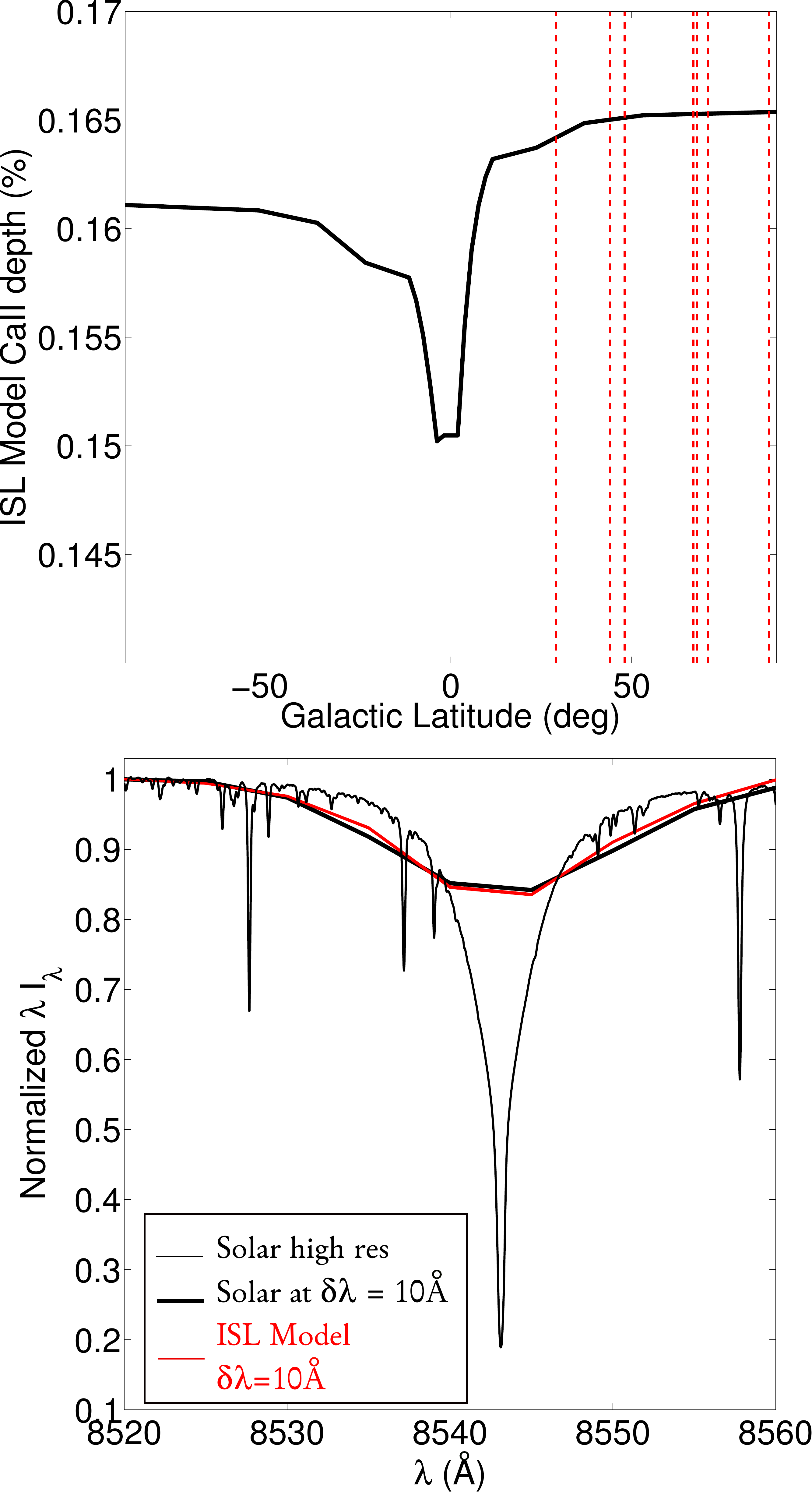}
\end{center}
\caption{{\it Top:} Depth of CaII in the \citet{Lehtinen13} ISL model as a function of Galactic latitude.  Locations of fields in this paper are shown as vertical red dashed lines. {\it Bottom:} Comparison of the \citet{Lehtinen13} ISL model and Solar spectra near the CaII line.  The Solar spectrum is shown both at its native resolution (thin black) and downgraded to match the ISL model at $\delta \lambda = 10$~\AA (thick black).}
\label{fig:islspec}
\end{figure}

The simulated NBS maps are then masked using an identical algorithm to the data, removing all pixels which appear in the simulated NBS map generated from the TRILEGAL based DSS $M_{AB} < M_{cut}$ catalog above $t$.  The FISL is taken to be a uniformly illuminating source with an amplitude set by the mean of the remaining pixels in the simulated TRILEGAL map.  

The final template ISL maps used in the fit are generated as the sum of the simulated DSS map and a uniform offset with an amplitude determined by the mean of the TRILEGAL FISL map mutliplied by the response function-convolved CaII absorption line with Solar depth (discussed in the following section).  These maps are shown in Figure~\ref{fig:isl}.

\subsubsection{CaII in the ISL}
\label{sS:islca2}

Since the ISL from the Milky way arises entirely from sources at z$\sim$ 0, spectral features, including the CaII line of interest will add coherently in the aggregate signal.  The ISL is composed of the entire stellar population, with a range of CaII depths, which could vary from location to location in the Galactic disk.  Unfortunately, a high spectral resolution ISL model in the appropriate region is not available in the current literature.  At moderate resolution, \citet{Lehtinen13} have compiled a global data-based model at a resolution of $\delta \lambda = 10$~\AA.  At this resolution (approximately 50$\%$ lower than the NBS), the CaII line is marginally resolved.  In Figure~\ref{fig:islspec}, we show the depth of the line as a function of Galactic latitude.  While a clear trend is present in the model close to the Galactic plane, for the fields included in this study (marked in red), the effect is negligible at $\sim 0.01\%$. Also shown in Figure~\ref{fig:islspec} is the high resolution Solar spectrum downgraded in resolution assuming a FWHM=$10$~\AA~Gaussian response function along side the average ISL spectrum from \citet{Lehtinen13}.  
As will be discussed in Section~\ref{S:sys},we take  the uncertainty in residual FISL amplitude from the TRILEGAL model to be $30\%$.  For purposes of the accuracy presented in this work, the 30$\%$ uncertainty dominates over the uncertainty in the ISL CaII depth. Additionally, we provide a test which allows us to bound the effect self-consistently using the NBS data.

\subsubsection{Test of the BISL CaII Line to Continuum}
\label{sS:islmodelcheck}

 While it is impossible to independently assess the total residual ISL from the NBS data as it is degenerate with the ZL signature, one can assess the accuracy of the BISL model through differential measurements. The star masking algorithm determines the level of residual ISL, subject to the choice of $M_{cut}$ and $t$.  
We define a test as follows and carry it out for a range of $M_{cut}$ and $t$ for all the target fields.
\begin{enumerate}
\item A star mask is generated using the algorithm described in \ref{sS:FISL} for an arbitrary $M_{cut}$ and $t$.
\item The mean intensity in the two dimensional NBS data is calculated for the pixels remaining after star and instrument masking.  The intensity below a masking threshold can be expressed as
\begin{multline}
\lambda I_{\lambda, Data, total}(M_{cut},t) =\\
 \lambda I_{\lambda, ISL}(M_{cut},t) + \sum\limits_{i} \lambda I_{\lambda, Data,i},
\label{eqn:dataavg}
\end{multline}
where i can be any other source of signal other than ISL (ZL, DC, DGL etc.) and thus independent of $M_{cut}$ and $t$.
\item Synthetic template images are made from TRILEGAL simulated catalogs in both the DSS-i2 and NBS filters assuming Solar CaII depth.  These images include stars down to $M_{AB}=$26. The appropriate NBS PSF is used for each field.
\item The mean in the simulated image is masked with the same values of $M_{cut}$ and $t$.
The quantity can be expressed as
\begin{multline}
\lambda I_{\lambda, Model, total}(M_{cut},t) =\\
 \lambda I_{\lambda, Model ISL}(M_{cut},t),
\label{eqn:modelavg}
\end{multline}
as there is only one component included.

\item Taking a derivative with respect to $M_{cut}$ and $t$ in Equations~\ref{eqn:dataavg} and \ref{eqn:modelavg} and equating the two, the expression 
\begin{multline}
\frac{d\lambda I_{\lambda, Data, total}}{dM_{cut}dt} =
\frac{d\lambda I_{\lambda, Model, ISL}}{dM_{cut}dt}
\label{eqn:derivavg}
\end{multline}
is valid in the limit that the model is a perfect description of the ISL.
\end{enumerate}

\begin{figure}
%\begin{center}
\includegraphics[width=0.5\textwidth]{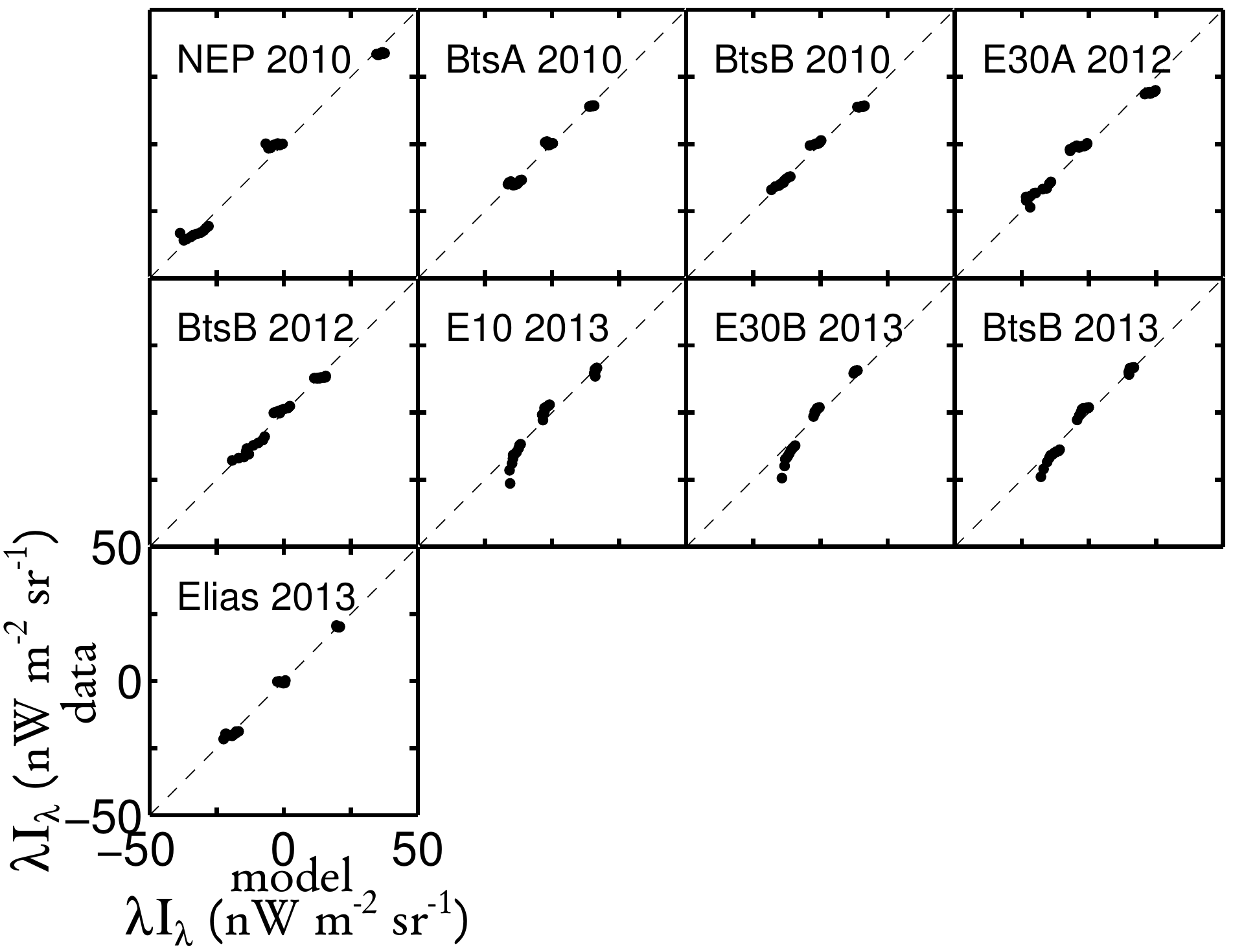}
%\end{center}
\caption{Stellar template and measurement consistency is tested in each field by varying the star mask and calculating the mean intensity in the remaining unmasked pixels.  The mask is generated using knowledge of the absolute calibration, PSF, and astrometry combined with fluxes and positions from the ancillary star catalog.  The model is generated from the TRILEGAL simulations and both assume CaII absorption at a Solar level. The dashed line shows unity correlation between data and model.  The best fit slope is $1.0 \pm 0.1$.}
\label{fig:islvcutmag}
\end{figure}

In Figure \ref{fig:islvcutmag} we show how the data and model in Equation~\ref{eqn:derivavg} relate.  Specifically, we plot the left hand side of Equation~\ref{eqn:dataavg} against the left side of Equation~\ref{eqn:modelavg} for the range $10 \leq M_{cut} \leq 12$ and $5$ $\leq t \leq 100 $~\nw.  Since each field has different offsets from the other components, we subtract a mean of each curve.  The dashed line is the case of the model being a perfect description of the ISL.  This test can allow us to probe errors included in the ISL model, the absolute calibration, the CaII absorption depth of the ISL, the astrometric solution and the extended PSF model. The measured slope of the correlation is $1.0 \pm 0.1$ averaged across all fields. One can interpret the relation as supporting the ISL having a solar CaII depth to better than $10\%$ accuracy.

The unity correlation builds good confidence in the quality of the BISL model, but does not help constrain the accuracy of the ISL models below the masking threshold. Ultimately, to quantify the uncertainty in our residual ISL amplitude and its effect on the ZL measurements, we rely on model accuracy tests in the literature, discussed in Section~\ref{S:sys}.

%%%%%%%%%%%%%%%%%%%%%%%%%%%%%%%%%%%%%%%%%%%
\subsection{Diffuse Galactic Light (DGL)}
\label{sS:dgl}

\begin{figure}
%\begin{center}
\includegraphics[width=0.5\textwidth]{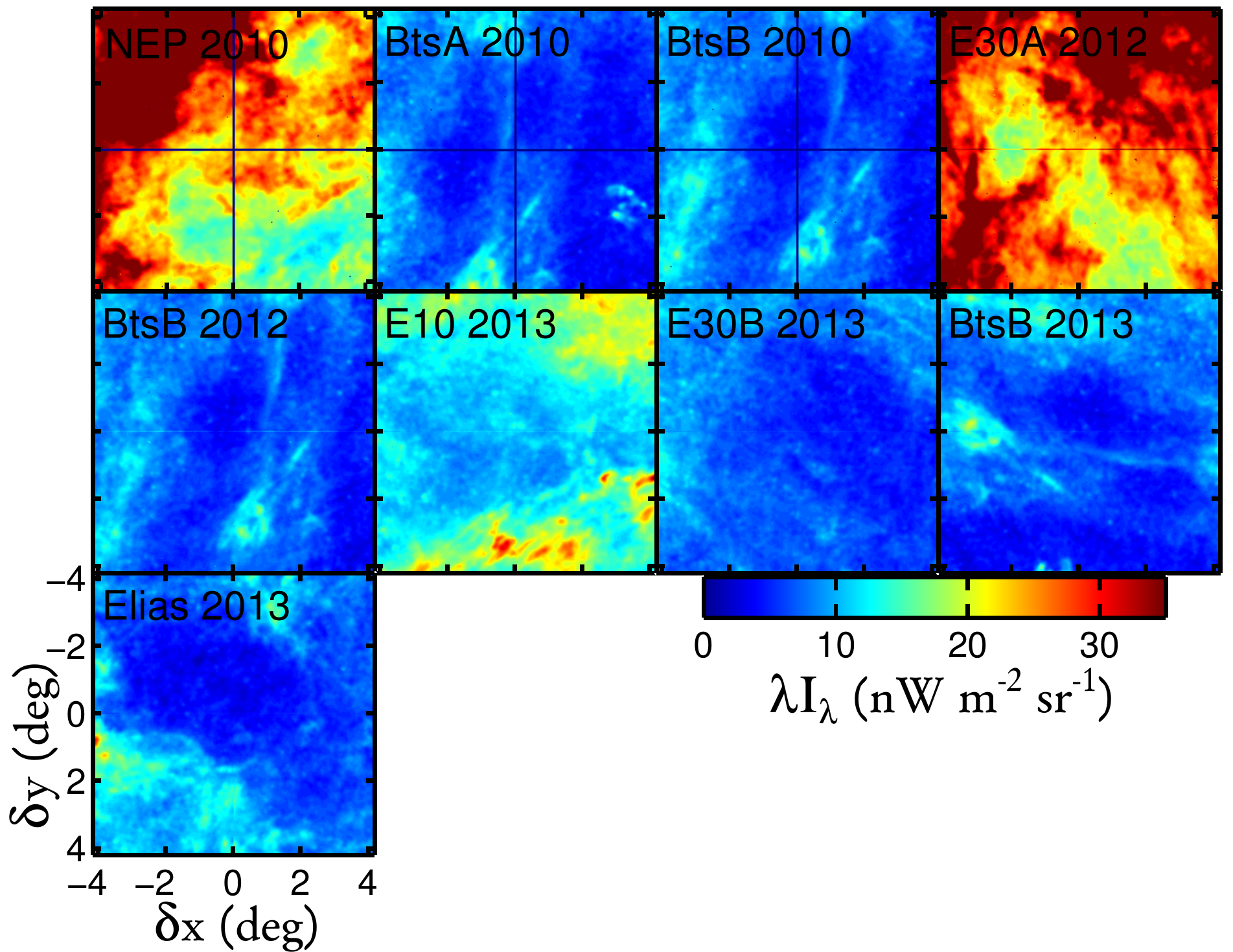}
%\end{center}
\caption{Template images of the DGL as it lands on the NBS array.  The images correspond to $G_{DGL,x,y}$ in Equation~\ref{eqn:2dmodel} and do not account for the Ca II absorption feature.}
\label{fig:dgl_grad}
\end{figure}

Diffuse Galactic Light arises from the same scattering phenomena as the ZL, only on larger scales.  In this case the illuminating light is the Galactic radiation field and the scattering medium is interstellar dust.  To estimate the continuum signal of DGL as it lands on the NBS FOV, for each field we interpolate the dust map of \citet{schlegeldust} on to the astrometrically registered grid for the pointing in each field.  It is then normalized to the appropriate continuum intensity using the scaling relation derived by \citet{araidgl}.  These maps, corresponding to $G_{DGL}$ in Equation~\ref{eqn:2dmodel} are shown in Figure~\ref{fig:dgl_grad}.  As is made evident by this figure, spurious alignment of intrinsic structure in the Galactic cirrus and the spectral response of each pixel on the array can produce a bias to the inferred ZL intensity from the CaII absorption feature.  By administering robust priors on the spatial distribution of the cirrus, this bias can be mitigated.

\begin{figure}
%\begin{center}
\includegraphics[width=0.5\textwidth]{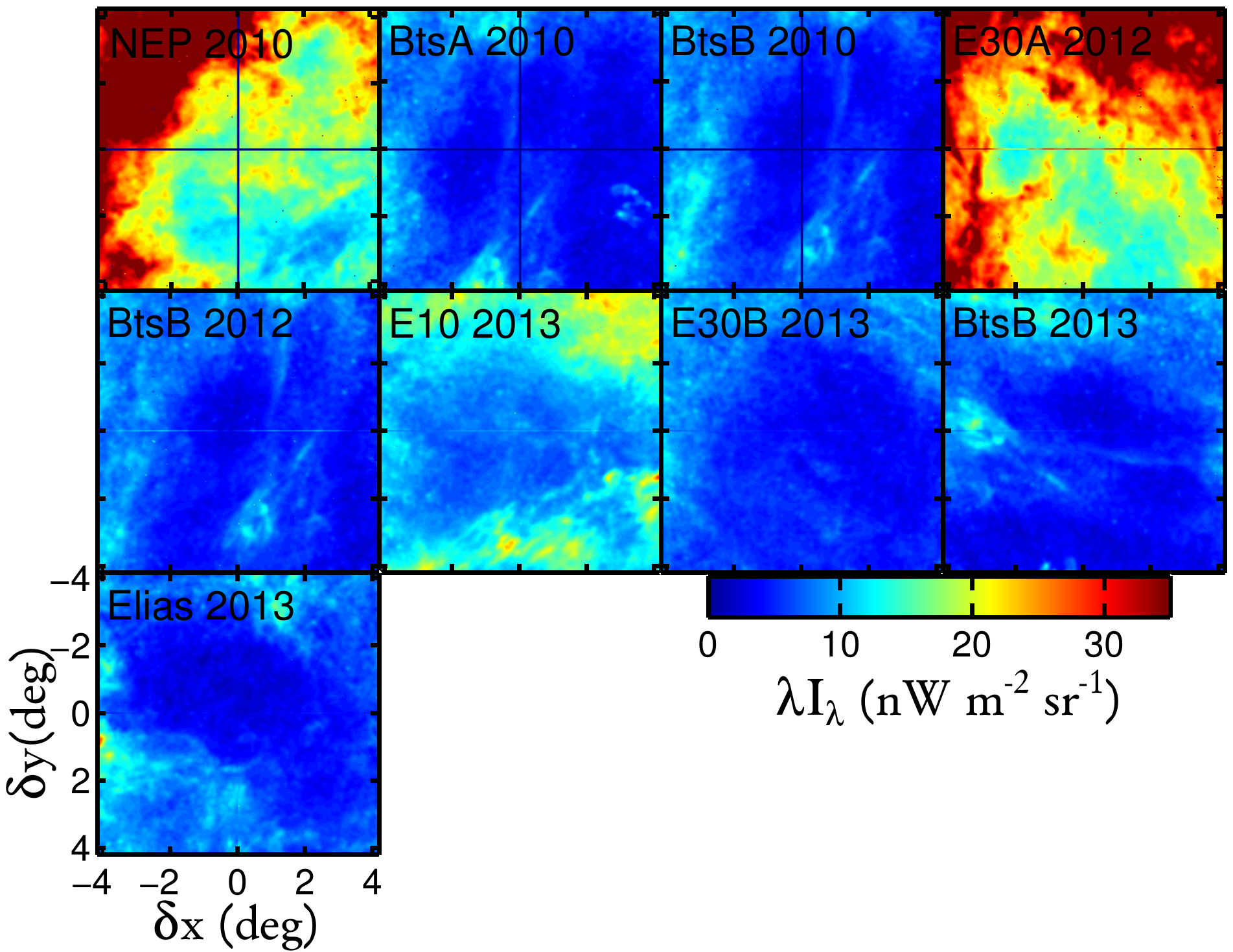}
%\end{center}
\caption{DGL distributions on the NBS FOV identical to Figure~\ref{fig:dgl_grad} after accounting for CaII absorption features in the DGL.  This quantity corresponds to $ G_{DGL,x,y} \int d\lambda \Lambda_{x,y} (\lambda) F_{\lambda,DGL}(\lambda)$.}
\label{fig:dgl_gradcaII}
\end{figure}

The case of DGL is further complicated by the fact that the interstellar radiation field also contains Fraunhofer lines, which are also scattered by the interstellar medium.  Figure~\ref{fig:dgl_gradcaII} shows the modelled DGL distribution after accounting for intrinsic absorption, it is simply modeled using the spatial DGL template in Figure~\ref{fig:dgl_grad} multiplied by the 2D solar absorption template for CaII. The model assumes Solar depth in the DGL CaII profile as it arises from a scattered ISL.

%%%%%%%%%%%%%%%%%%%%%%%%%%%%%%%%%%%%%%%%%%%%%%%
\subsection{Zodiacal Light (ZL)}
\label{sS:zl}

\begin{figure}
%\begin{center}
\includegraphics[width=0.5\textwidth]{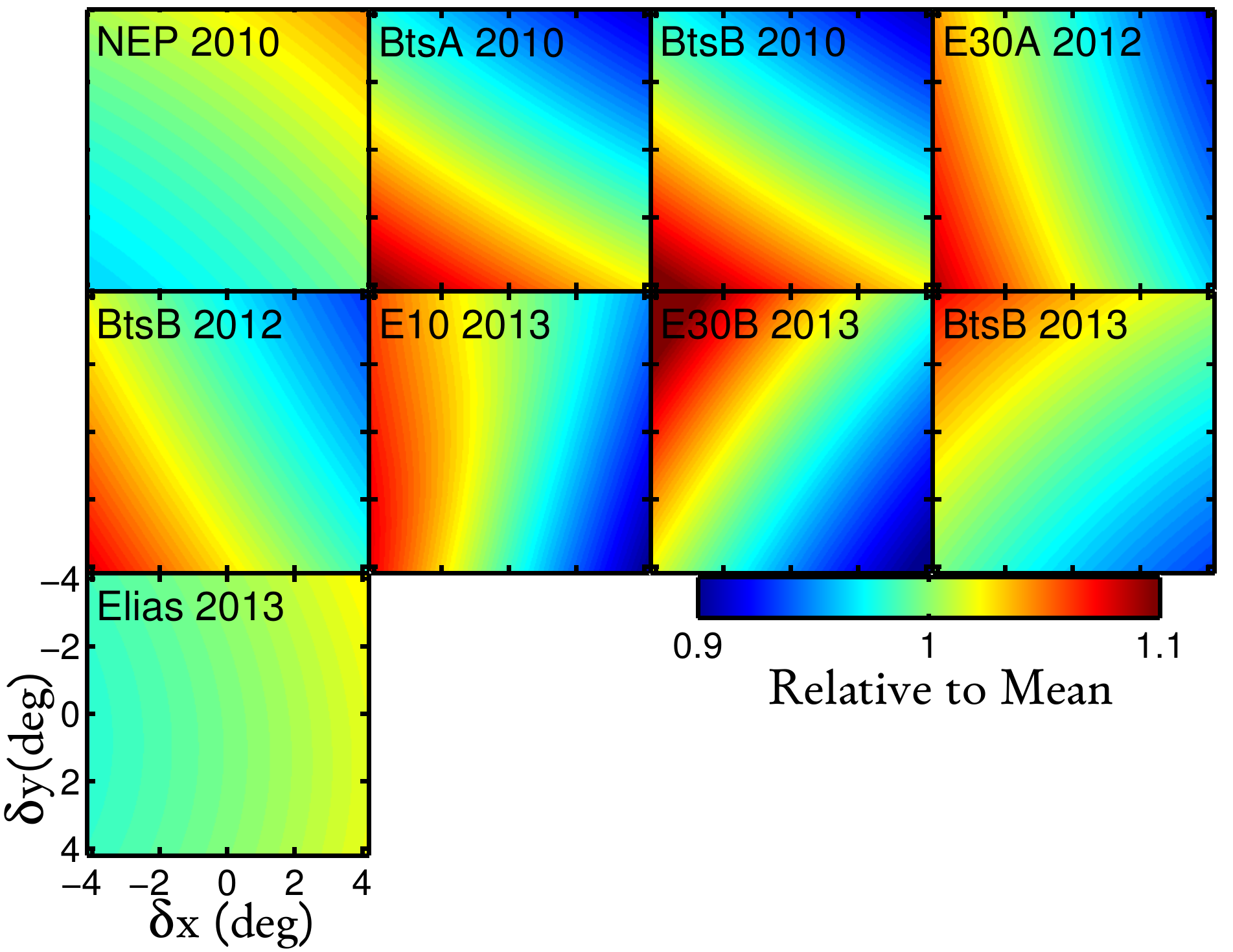}
%\end{center}
\caption{The gradient in ZL intensity across the NBS field of view as described by the \citet{Kelsall98} model at 1.25~$\mu$m. This is representative of the continuum only and does not account for the wavelength distribution of the NBS.  All fields are normalized to their mean and shown on the same color-scale.}
\label{fig:zodi_grad}
\end{figure}

\begin{figure}
\includegraphics[width=0.5\textwidth]{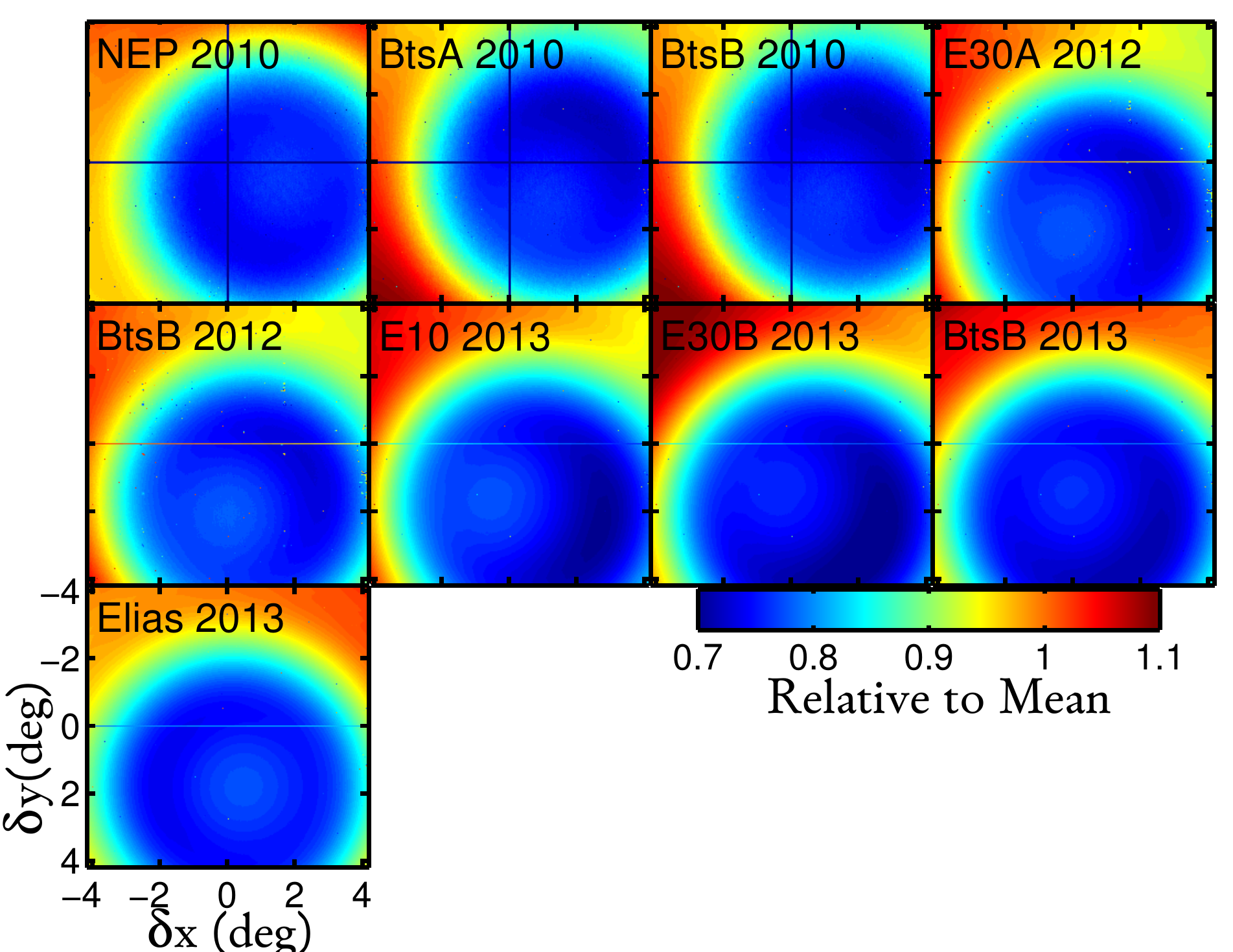}
\caption{ZL spatial and spectral templates for each field as observed by the NBS. They are generated by multiplying the gradient shown in Figure~\ref{fig:zodi_grad} by the CaII absorption line profile for a uniform illumination.  All fields are normalized by the mean.}
\label{fig:zodi_grad_caii}
\end{figure}

Because of the $8.5^{\circ} \times 8.5^{\circ}$ instantaneous FOV of the NBS, and the desired level of accuracy, assuming a uniform ZL intensity can produce a $\sim5\%$ error in the derived ZL amplitude. We use prior models of the spatial distribution of the ZL to reduce error from spatial gradients.  Figure~\ref{fig:zodi_grad} shows ZL gradients computed from the \citet{Kelsall98} model over the NBS FOV for each field.  These gradients are normalized by the mean over the FOV.  Figure~\ref{fig:zodi_grad_caii} shows the effect these gradients have on the ideal CaII absorption feature as shown on the Solar spectrum in \citet{Korngut13}.

%%%%%%%%%%%%%%%%%%%%%%%%%%%%%%%%%%%%%%%%%%%%%%%
\section{Model Fits}
\label{S:fits}

After considering all of the effects discussed in Section~\ref{S:2D}, the models are fit to the data in each field using the algorithm described below.  Before fitting, the NBS images are processed following the procedures described in Section~\ref{S:data}.  We fit the processed images as follows.

\begin{enumerate}
\item We generate a two dimensional template consisting of all fixed components including the DGL, BISL and FISL.  There are no free parameters in these products, as everything is generated using ancillary data and empirically determined instrument characteristics.

\item We add a Zodiacal Light template to the model.  This consists of the ideal CaII absorption profile \citep{Korngut13} mapped to the NBS wavelength response after accounting for the gradient across the FOV determined by the ancillary ZL model normalized by the mean.  The ZL template has a free normalization $A_{ZL}$.  

\item We add a free spectrally flat offset $C$ to account for the EBL.

\item Both the data and model are masked using the pixel masking criteria described in Section~\ref{S:data} and the stellar masking described in Section~\ref{sS:isl}.

\item One dimensional spectra are extracted for both the model and data and the reduced $\chi^2$ statistic is computed.  The statistical errors on the one dimensional spectra are computed from the RMS variation within an iso-wavelength bin divided by the square root of the number of pixels used.
\end{enumerate}

\begin{figure}[t!]
\begin{center}
\includegraphics[width=0.95\textwidth]{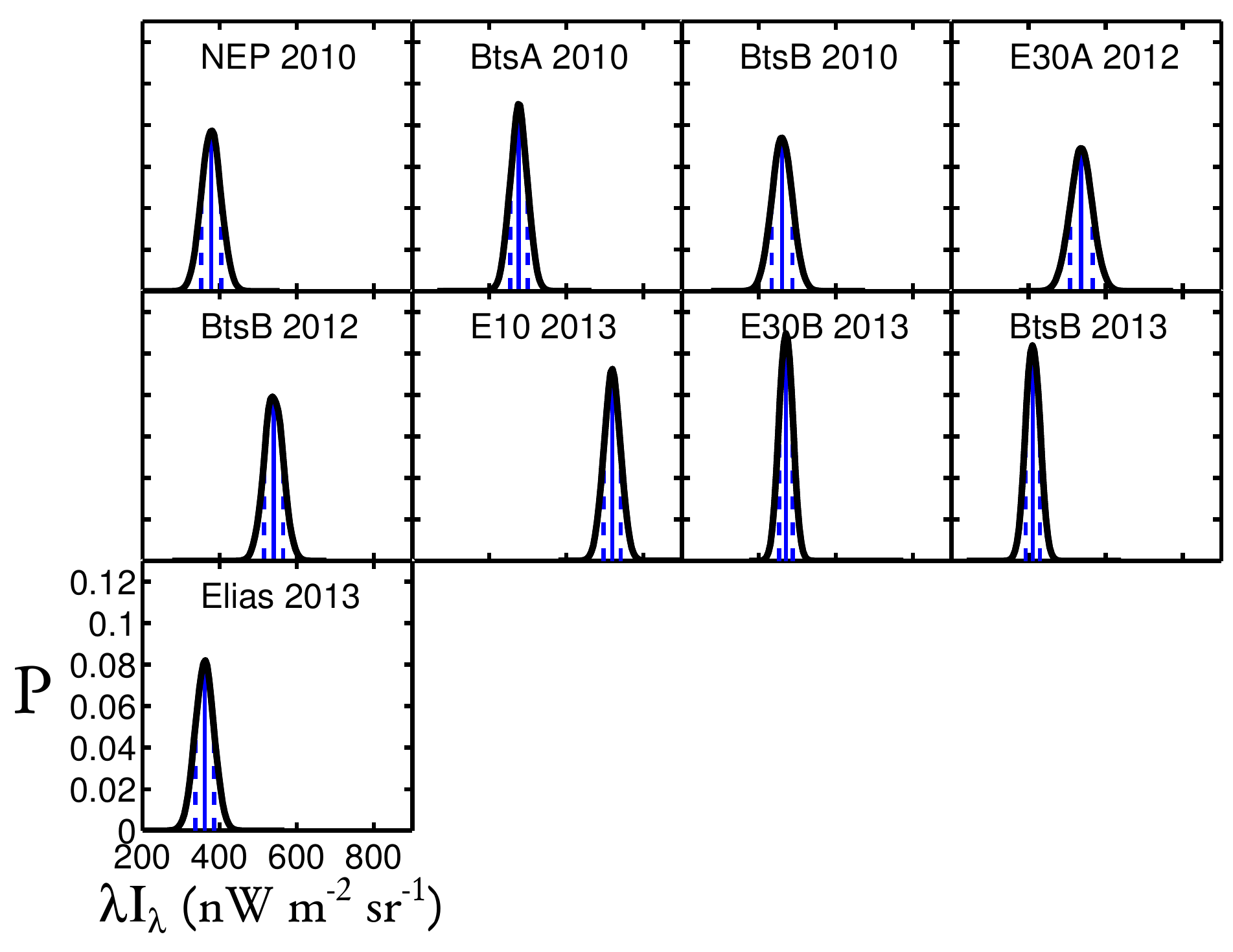}
\end{center}
\caption{Probability density functions for the ZL continuum intensity parameter $A_{ZL}$ at 8520~\AA $ $ after marginalizing over the spectrally smooth offset parameter $C$. Solid blue lines show the mean with the two dashed lines $\pm1\sigma$.  These distributions represent statistical sources of error only.}
\label{fig:liklihood}
\end{figure}

We vary and fit the ZL normalization parameter $A_{ZL}$ and the offset $C$.  After the two dimensional $\chi^2$
distribution is generated, the offset parameter is marginalized over to produce the probability density functions (PDFs) shown in Figure~\ref{fig:liklihood}.  The PDFs are well behaved with single prominent peaks.  The 1$\sigma$ statistical error bars presented in this paper are taken from the width of these PDFs, which have typical values of $\sim$23 nW~m$^{-2}$~sr$^{-1}$.  We note that the level of statistical uncertainty is entirely consistent with the values presented in \citet{Korngut13} which were determined by Monte-Carlo simulations using realizations of measured laboratory noise.  The measured spectra in 1D are shown along with the best fit models in Figure 13.  The error bars in the 1D spectra are determined by the RMS of the pixels in an iso-wavelength bin divided by the square root of the number of pixels.

\section{Systematic Errors}
\label{S:sys}

In addition to the statistical errors determined in Section~\ref{S:fits}, a number of additional systematics must be accounted for.

\subsection{Instrument Systematics}

\subsubsection{Spectral Correction}
The shortest wavelength probed by the DIRBE instrument, upon which the Kelsall and Wright models are based, is 12500~\AA.  Therefore, to compare the inferred NBS continuum measurements at 8520 ~\AA $ $ with the ancillary models, a spectral correction is needed.  To do this, we rely on $\frac{\lambda}{\Delta\lambda}\sim 20$ measurements of the ZL continuum spectrum made by the CIBER LRS.  As the LRS is co-mounted with the NBS on the rocket, the template is generated using the same fields observed at the same epochs.  Construction of the template is given in \citet{MatsuuraLRS}.  The measured ratio between the two wavelengths for ZL is $\frac{\lambda I_{\lambda,ZL} (\lambda = 8520\rm{\AA})}{\lambda I_{\lambda,ZL} (\lambda = 12500\rm{\AA})} = 1.25 \pm 0.1 $.  It should be noted that this ratio is substantially smaller for the Solar spectrum $\frac{\lambda I_{\lambda,Solar} (\lambda = 8542\rm{\AA})}{\lambda I_{\lambda,Solar} (\lambda = 12500\rm{\AA})} = 1.5$.  This discrepancy is accounted for by spectral reddening from the IPD \citep{akarizl}.  The inferred NBS values for each field are divided by this factor when comparing to DIRBE model predictions.
\subsubsection{Calibration Error}
The absolute calibration factor used to convert an observed photocurrent to a physical intensity relies on the laboratory measured conversion (see section\ref{S:data}).  Laboratory measurements have shown reproducibility within $2.5\%$.
\subsubsection{Dark Current}
To quantify the error introduced by DC subtraction, we compare the low-noise rail DC template image to the in-flight image taken with the cold shutter closed during the 2013 flight. This represents a worst-case condition, as the focal plane temperature stability during this flight was substantially worse than the previous flights.  The instrument included an active temperature control system that regulated the focal plane temperature.  During the 2009 and 2012 flights, the system worked nominally and the focal plane was regulated within an rms of $\pm 15\mu$K.  Due to the base temperature in the 2013 flight operating outside the dynamic range of the control unit, the 2013 flight was conducted with only passive thermal stability.  A pixel wise correlation of the DC rail template and flight measurement show a linear trend with a slope of 0.4.  We pass a residual map of the DC template after subtracting a scaled down image through the CaII fitting pipeline to quantify the amplitude of a systematic error.  This is taken to apply randomly to all fields with an amplitude of $1\sigma = \pm 17$ \nw, referred to the ZL continuum intensity at 8520~\AA. 

\subsubsection{Flat Field}
 The accuracy of the flat field correction is determined by fitting a CaII line response to the ratio of flat fields obtained at different intensity levels in the laboratory.  Using this technique, we estimate the error introduced by the flat field to be a multiplicative factor of 0.3$\%$, far sub-dominant to the other multiplicative errors such as the calibration factor and the spectral correction.  This uncertainty level is consistent with an estimate produced by measuring the CaII signal in Solar light coupled to the laboratory by fiber and comparing to precise archival Solar spectra \citep{Korngut13}.

\subsection{Modeled Astrophysical Systematics}
\subsubsection{DGL}
The scaling of the \citet{schlegeldust} maps to the NIR (which sets $A_{DGL}$) implemented in this analysis is set by the empirically determined relation of \citet{araidgl} from CIBER LRS data.  
The relation has a $30\%$ error associated with it, which directly propagates to the modelled DGL component.

\subsubsection{Bright ISL}
As discussed in detail in Section~\ref{sS:islmodelcheck}, the parameterization of the DSS-based ISL model with $M_{cut}$ and $t$ encompasses the accuracy of the ISL model along with the measured instrument parameters which go into the generation of synthetic star maps such as the extended PSF, the distortion field and astrometric solution.  For a perfect model with zero systematic error, a measurement of the ZL intensity should be constant with any choice of $M_{cut}$ and $t$.  For less aggressive mask cuts, the ISL makes up a higher fraction of the total signal. For very aggressive cuts, too few pixels remain for spectral extraction.  

%%%%%%%
\onecolumn
\begin{figure}
\begin{center}
\includegraphics[width=.95\textwidth]{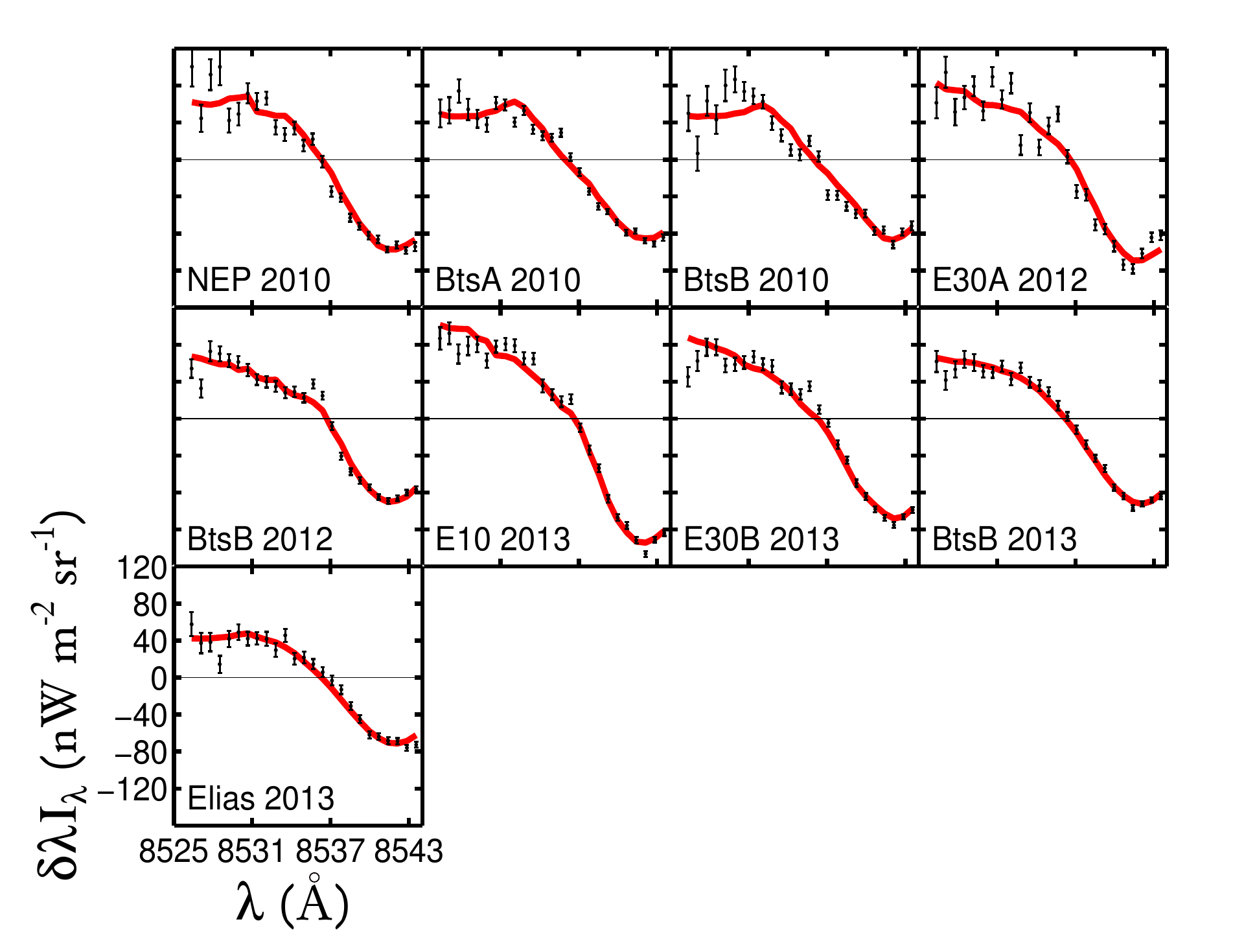}
\caption{One-dimensional measured spectra ({\it black data points}) with best fit models ({\it red curves}) for all fields. To show the variation in line depth, all axes are matched and a mean has been subtracted from all spectra.  The models are fit with only a ZL amplitude and constant offset as free parameters.  The variation in line profile between fields is due to structure in the DGL and the gradient in ZL at each location, constrained entirely by external information.  The DGL and ZL gradients are included in the red curves.}
\end{center}
\label{fig:specfit}
\end{figure}
%%%%%%%%%%%%%%%%%%%%%%%%%%%%%%%%%%%%%%%%%%%%%%%
\twocolumn

\subsubsection{Faint ISL}
The amplitude of the ISL at magnitudes fainter than the masking threshold is based on the Trilegal star count model and an assumed Solar CaII absorption depth.  Comparisons of the output of this model and numerous source count measurements agree to within $\pm30\%$ \citep{trilegal}.  This uncertainty is propagated directly to our measurements.

\subsection{Error Propagation}
Because the chosen celestial field locations span a large range of Galactic and Ecliptic latitudes, the component breakdown of errors is unique for each target. Other systematic errors, such as the instrumental calibration factor, affect our measurements multiplicatively and systematically push all field measurements up and down together.  It is therefore necessary to understand the interplay between the various effects on a field-by-field basis. Consequently, we calculate the combined systematic and statistical errors uniquely in each field. 

\begin{figure*}
\begin{center}
\includegraphics[width=1.0\textwidth]{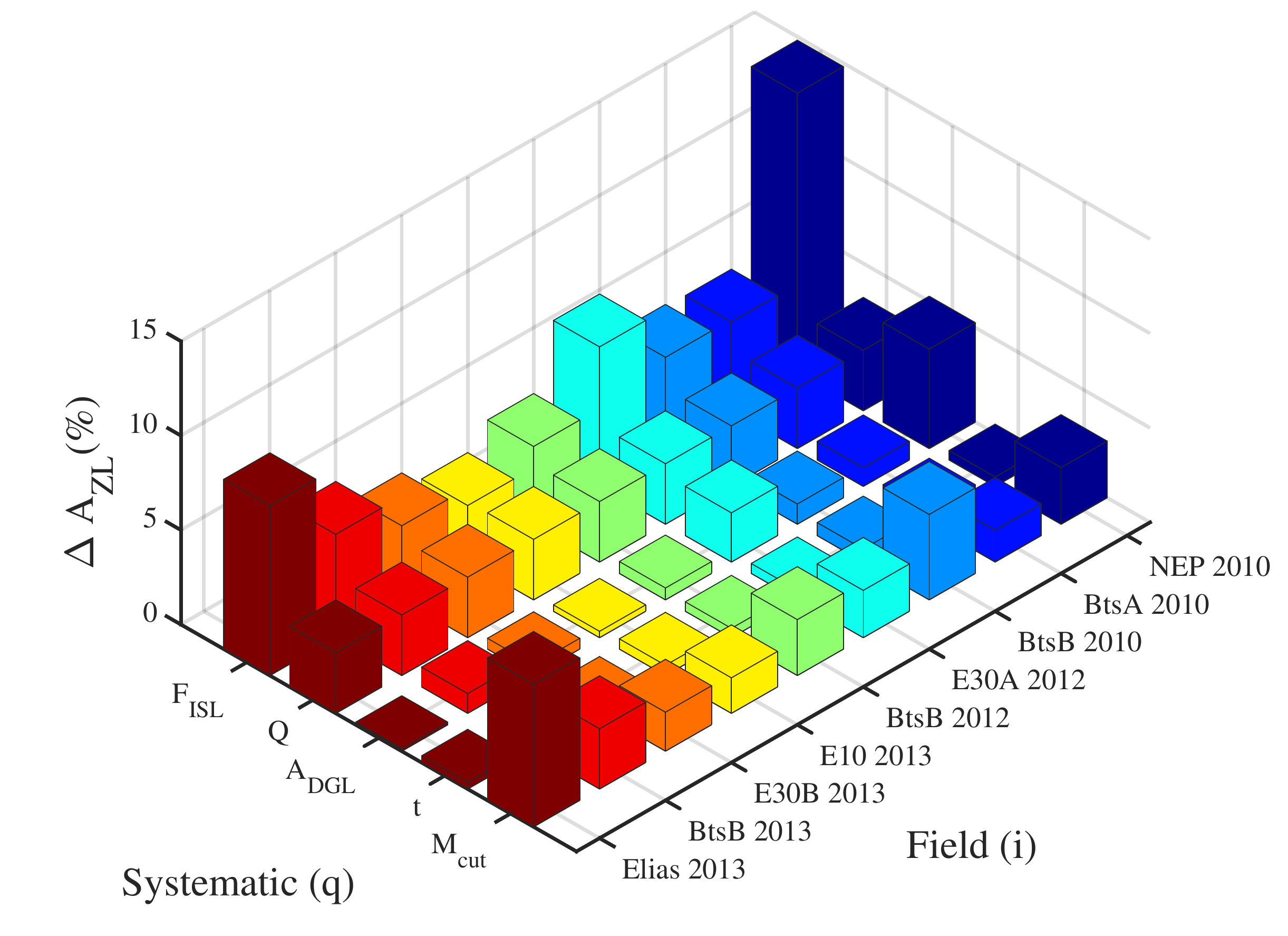}
\caption{Variation in the estimate of $A_{ZL}$ attributed to individual systematic errors for each field.  For simplicity, we define the parameter $Q$ to be the product of the calibration and wavelength extrapolation systematics.  Due to the variation in stellar properties, dust content and ZL amplitude in each location, the breakdown of systematic contributions is unique for each field.}
\label{figure:sysbar}
\end{center}
\end{figure*}

To quantitatively illustrate the complex interaction of the systematic errors in each field, it is useful to define a parameter
\begin{equation}
\Delta A_{ZL_{q,i}} = \frac{A_{ZL_{q_{max},i}} - A_{ZL_{q_{min},i}}}{2A_{ZL_{q_{nominal},i}}},
\end{equation}
where $q$ represents a given systematic error contribution and $i$ represents the target field.  To compute $\Delta A_{ZL_{q,i}}$, we fix all systematic errors other than $q$ to their a-priori assumed nominal values, and calculate $A_{ZL_{q_{max},i}}$ and $A_{ZL_{q_{min},i}}$, which are the best fit ZL amplitudes under the assumptions that systematic $q$ is at its maximum and minimum value within an allowed range respectively.  Figure~\ref{figure:sysbar} displays $\Delta A_{ZL_{q,i}}$ for each value of $q$ and $i$, where the allowed ranges of systematics are given in Table~\ref{table:sys}.

\begin{table}
\begin{tabular}{c|c|c|c}
\hline
\hline
Parameter & low & high & units \\
\hline
$M_{cut}$ & 12 & 10 & $m_{AB}$ \\
$t$ & 5 & 100 & \nw \\
$\delta A_{DGL}$ & 0.7 & 1.3 & -- \\
$\delta A_{FISL}$ & 0.7 & 1.3 & -- \\
Cal & 618 & 650 &  \nw / \\ 
 &  &  & (e-/s)  \\ 
$\frac{\lambda I_{\lambda,ZL} (\lambda = 8520\rm{\AA})}{\lambda I_{\lambda,ZL} (\lambda = 12500\rm{\AA})}$ & 1.24 & 1.26 & --\\
FF & 0.997 & 1.003 & --\\
\end{tabular}
\caption{Parameter ranges explored in the systematic error analysis.  For all parameters, a uniform distribution spanned by this range is explored.\label{table:sys}}
\end{table}

\begin{figure}
\begin{center}
\includegraphics[width=0.5\textwidth]{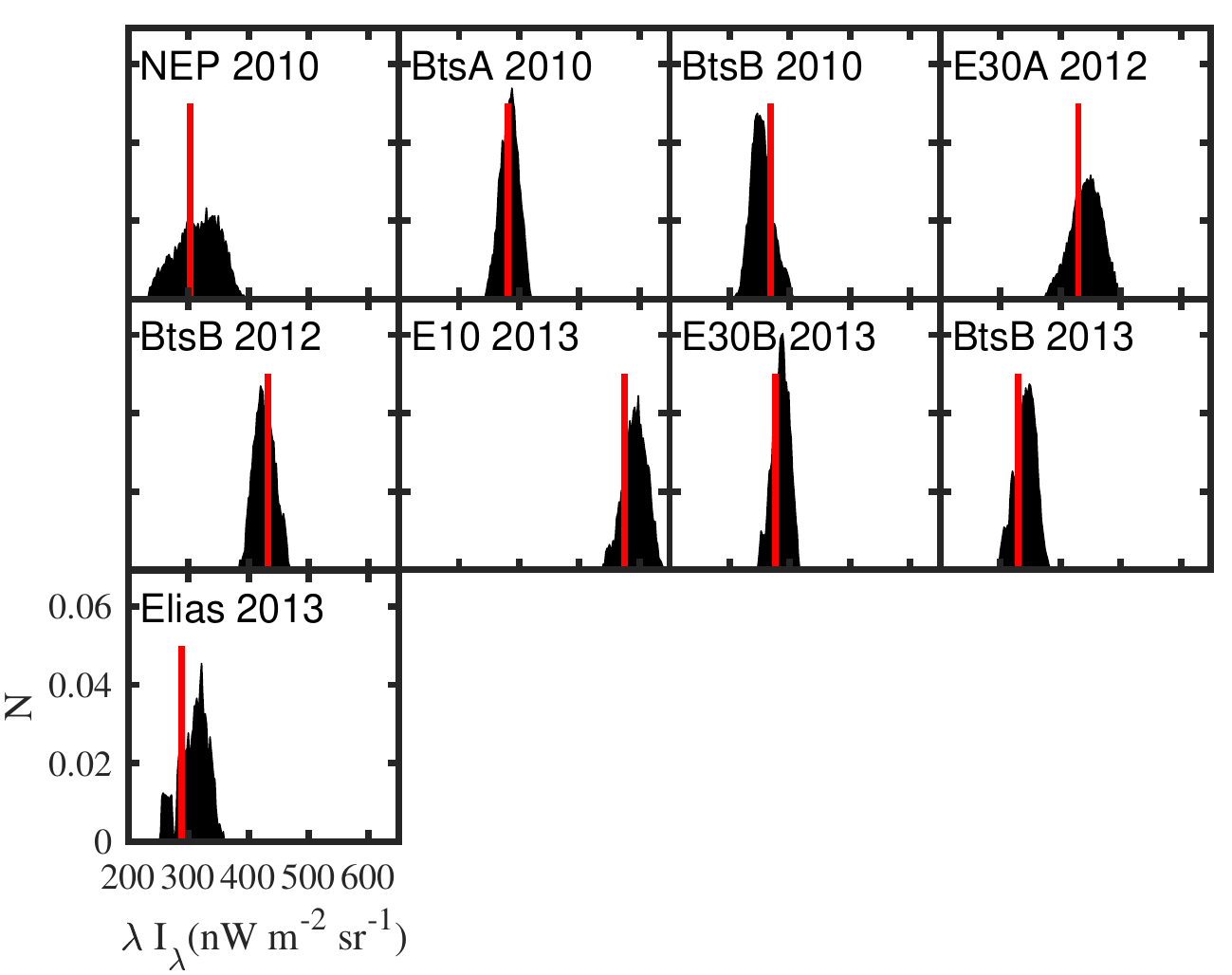}
\caption{Histograms of $A_{ZL}$ for each field's allowed parameter space of systematic errors.  The histograms are normalized by the number of permutations explored.  The red vertical line denotes the a priori assumed nominal set of parameters, which is different from the most visited value in some cases. }
\label{fig:sysfields}
\end{center}
\end{figure}

While $\Delta A_{ZL}$ is useful in illustrating the extreme values allowed by isolated effects, to obtain a full understanding of allowed values of $A_{ZL}$ it is necessary to carry out the entire analysis described in Section~\ref{S:fits} under every permutation of systematic error parameters.  To do this, we explore a grid of 2505 discrete permutations of the allowed error contributions, calculating $A_{ZL}$ at each point for each field.  The histograms of all outcomes of $A_{ZL}$ are shown in Figure~\ref{fig:sysfields} after normalizing by the total number of permutations.

The varied shapes of the distributions reflect the characteristics of each field.  In locations that have a large contribution of DGL and ISL compared to the total signal such as NEP, the distribution is wide and asymmetric.  Regions that are dominated by ZL, such as Elat 30B, have well defined peaks in the distribution.  

%%%%%%%%%%%%%%%%%%%%%%%%%%%%%%%%%%%%%%%%%%%%%%%
\section{Model Comparisons}
\label{S:models}

As discussed in the introduction, nearly all NIR EBL absolute spectro-photometry measurements 
in the literature rely on ZL foreground subtraction based on models generated from geometrical fits to the DIRBE data.  In particular, measurements that report a higher level of EBL were generated using the model of \citet{Kelsall98} and fainter estimates rely on \citet{Wright2001}. Because these two models bookend the range of reported EBL, we concentrate our analysis on applying NBS data as a test to the models and henceforth, a probe of whether a brighter or fainter EBL is favored.

\begin{figure*}
\begin{center}
\includegraphics[width=0.52\textwidth]{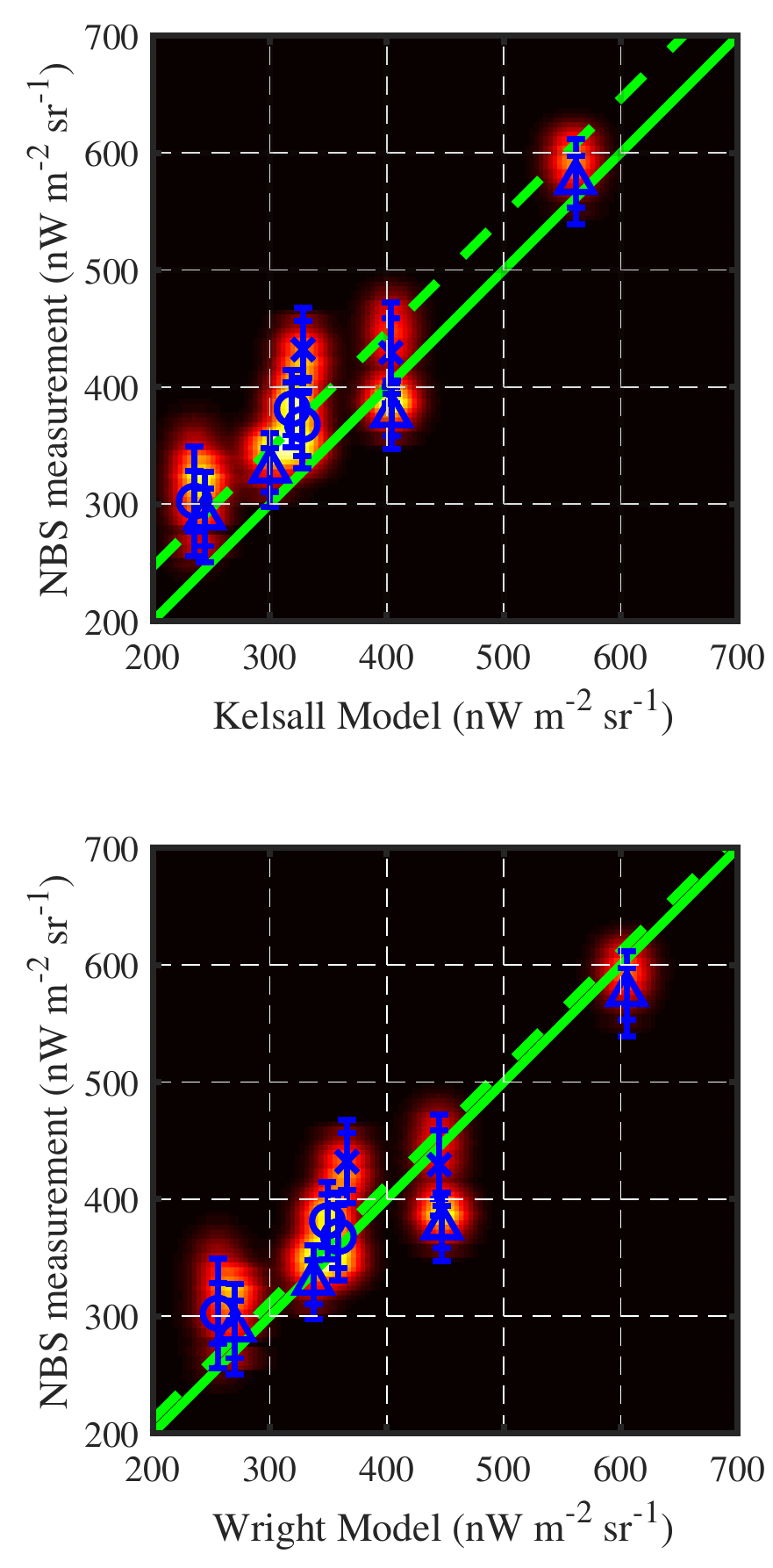}
\end{center}
\caption{Comparison of the NBS ZL intensity measurements to the Kelsall (top) and Wright (bottom) models at 12,500~\AA. The colorscale errors are a two dimensional visualization of the distributions given at each field from the systematic error analysis.  Blue points show the amplitude estimates assuming nominal values for all systematic errors.  There two sets of blue error bars which represent the statistical only (inner), and combined statistical and systematic (outer). The blue marker type denotes the rocket flight in which the measurement was performed (2010 are circles, 2012 are xs and 2013 are triangles). The solid green line shows a unity correlation and the dashed green shows the unity correlation plus a best fit offset for each model. }
\label{fig:modelcomp}
\end{figure*}

In Figure~\ref{fig:modelcomp} we show the correlation of the absolute ZL intensity inferred by the NBS to the Kelsall and Wright models after extrapolation to 12500~\AA. In this plot, the blue points convey the estimates generated under nominal systematic error assumptions  The color scale encompasses the regions in which the blue points move around under the allowed range of uncertainty for all of the systematic errors.  The vertical distribution of the colorscale directly corresponds to the shapes of the histograms in Figure~\ref{fig:sysfields} and the horizontal distribution is assumed to have a Gaussian error $\sigma=15$~\nw, the error reported by \citet{Kelsall98}. 

\begin{figure}
\begin{center}
\includegraphics[width=0.5\textwidth]{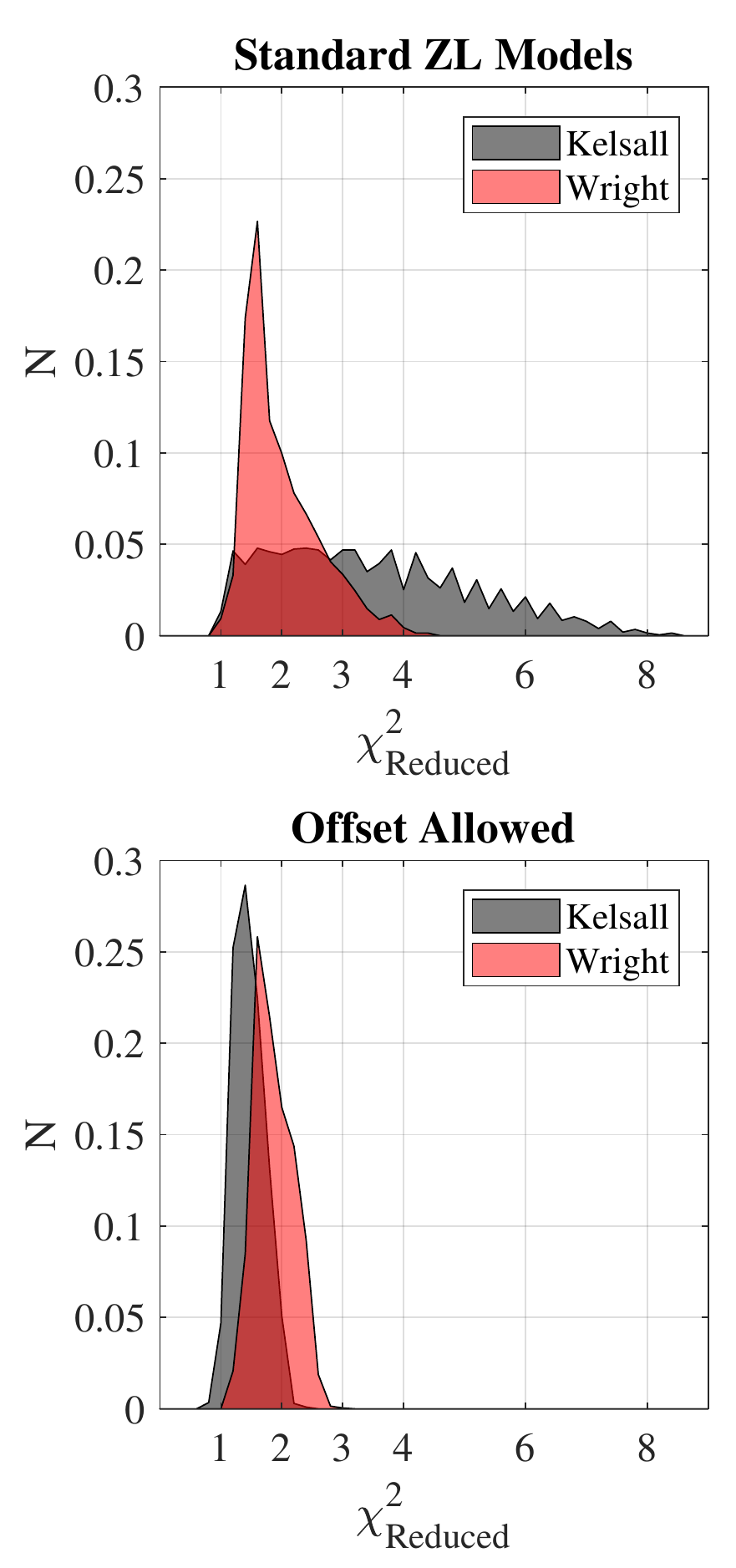}
\caption{Reduced $\chi^{2}$ distributions combining statistical and systematic errors as a test of goodness of fit to both the Kelsall (black) and Wright (red) models. {\it Top:} Evaluating $A_{ZL}$ compared to the two ZL models. {\it Bottom:} After including an additional degree of freedom represented by a best fit offset.}
\label{fig:chi2}
\end{center}
\end{figure}

To assess how well each model describes the NBS measurement, we calculate a reduced $\chi^{2}$ statistic at every explored location in parameter space. As shown in Figure~\ref{fig:chi2}, when no additional free parameters are allowed in the models, the $\chi^{2}$ distributions look very different for the two cases. 

\begin{figure}
\begin{center}
\includegraphics[width=.5\textwidth]{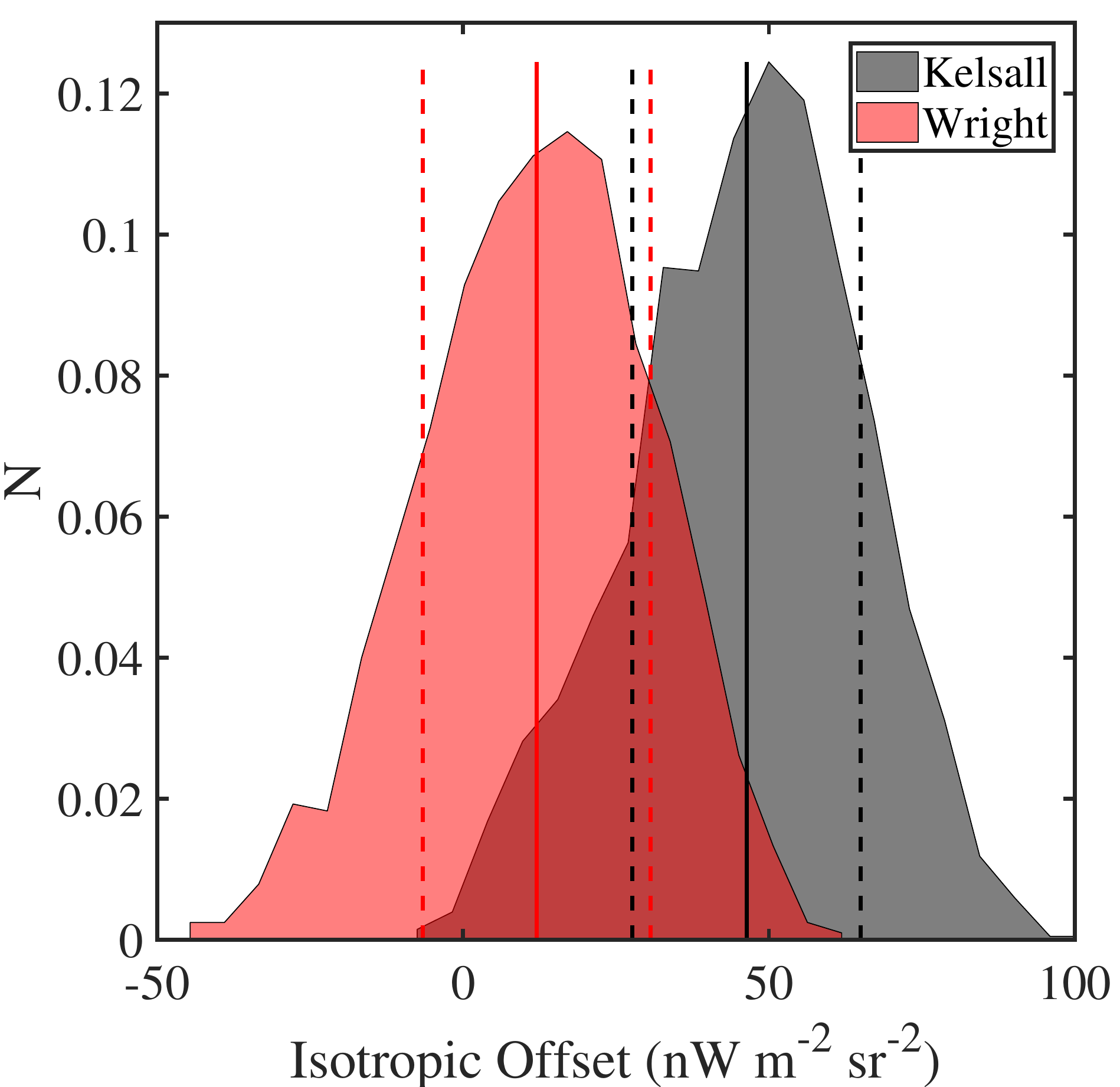}
\caption{Distributions of inferred 12,500~\AA~values for a ZL offset for the Kelsall (black) and Wright (red) model predictions.  The mean and $1\sigma$ range are denoted with the solid and dashed lines respectively. The Wright model is consistent with zero to within $1\sigma$ and the Kelsall model shows an offset $2.4\sigma$ above zero. \label{fig:meanoff}}
\end{center}
\end{figure}

In the case of the Kelsall model, the majority of the data points lie systematically higher than the model predictions, leading to a reduced $\chi^{2}$ distribution that is very broad with a tail that extends all the way up to a value of 8 and a mean value of 3.5.  For the Wright model, the distribution is more symmetrical with a mean value of 2.0.  In this case, the data points are distributed both above and below the model predictions and the goodness of fit is limited by field-to-field scatter.

Also shown in Figure~\ref{fig:chi2} are $\chi^2$ distributions calculated under the simplest modification to the foreground model; the addition of a single free parameter in the form of a constant ZL offset. When this additional free parameter is included in computing $\chi^{2}$, the distribution of the modified Kelsall model changes dramatically.  It displays a sharp symmetrical peak with a mean value of 1.5, coming down even after accounting for the additional degree of freedom.  The Wright model distribution in this case is more symmetrical than without an offset, and the mean value reduces slightly to 1.9.  The histograms of the best fit offset under all permutations of systematic errors for both the Kelsall and Wright models are given in Figure~\ref{fig:meanoff}.  The data suggest an offset from the Kelsall model with an amplitude of $46 \pm 19$~\nw at 12500~\AA.  The Wright model is consistent with zero, with a most likely value of $12 \pm 19$~\nw at 12500~\AA.

%%%%%%%%%%%%%%%%%%%%%%%%%%%%%%%%%%%%%%%%%%%%%%%
\section{Discussion}
\label{S:discuss}

\subsection{Model testing and EBL implications}

We present new measurements of the ZL absolute intensity in the NIR through Fraunhofer absorption line spectroscopy. Through these measurements, we provide a test of the two ZL models most heavily cited in absolute NIR spectro-photometric measurements. After accounting for the interaction between statistical and systematic errors, we find the data favor an absolute ZL intensity that is somewhat brighter than predicted by the Kelsall model.  The total observed intensity is closer to the Wright model, but with additional field-to-field scatter. 

 The offset distributions in Figure~\ref{fig:meanoff} can be loosely interpreted as residual ZL that would be falsely interpreted as EBL in an absolute measurement. The Wright model distribution has a mean consistent with zero within $1\sigma$, whereas the Kelsall model distribution is $2.4\sigma$ above zero.  The widths of these distributions limit the confidence with which we can rule out inferred EBL amplitudes in the literature, but a brighter-ZL, weaker-EBL interpretation is favored by our data at modest significance.
 
\subsection{Evidence for an Additional IPD component?}
We introduced a single free parameter that posits the addition of a ZL offset.  The DIRBE experiment, on which the Kelsall model is based, was a NASA mission operated in low Earth orbit. The model was generated by fitting a geometrical parameterization to the annual modulation in DIRBE-measured intensity.  This natural variation arises from the change in line of sight through the inclined IPD cloud as the Earth orbits around the Sun and through the cloud.  Because all data constraints are derived from a differential signal, the Kelsall model is by design insensitive to any isotropic signatures, which contribute intensity that does not vary with an annual modulation as observed from 1 AU.

The Wright model was designed to include all observed flux, under the assumption that the total sky brightness at 25~$\mu$m, where the ZL peaks, was entirely from the ZL.  The shorter wavelength intensities were then predicted by spectral extrapolation using the measured ZL color.  Therefore, the Wright model would include flux from an isotropic component at the cost of potentially attributing some EBL flux at 25~$\mu$m to the ZL.  

The idea of an isotropic ZL component which would evade detection in  geometrical studies has been posited by several studies in the literature. \citet{chary10}, in their investigation into FIR background sources suggested their data could be explained by the existence of thermal emission from a $ 53\pm 16$~K diffuse source around the outer solar system ($>~$200 AU).  However, \citet{tsumura18} argue through a series of model fits to the thermal emission that the NIR flux contribution of such a feature must be exceedingly small, on the scale of 1~\nw at 12500~\AA,~ which would be undetectable in this study. The calculations in \citet{tsumura18} are limited to ZL components in the outer solar system, and do not constrain components at 1~AU from the Sun. 

Recent dynamical simulations of the IPD such as those presented in \citet{poppe16} and \citet{nesvorny10} predict a heliocentric isotropic IPD distribution in the inner solar system supplied by debris from long-period Oort-cloud comets (OCC) dynamically mixed in orbital space. \citet{nesvorny10} found their model fits to mid-infrared IRAS data were dramatically improved when including this OCC component containing $\sim 5\%$ of the IPD residing in an isotropic cloud in the inner solar system. The Kelsall geometrical model contains a smooth cloud, a series of three bands, a solar ring and an Earth-trailing blob, but does not contain a component resembling the OCC posited by \citet{poppe16} and \citet{nesvorny10}.

\citet{sano20} presented a re-analysis of DIRBE data which examined the sky brightness modulations as a function of solar elongation.  Including dependence on the angle of the scattering function led them to the conclusion that a spheroidal ZL component in the inner solar system that was not included in the Kelsall model improves consistency with the data.  They fit two models, with the brighter suggesting an isotropic ZL amplitude as bright as $19.45 \pm 1.99$~\nw at 12500~\AA~attributed to the OCC component. 
 
Our findings based on 9 fields observed over 3 flights support a similar hypothesis, with an additional $46\pm 19$~\nw of ZL extrapolated to 12500~\AA.  However, we cannot meaningfully test for isotropy.  The OCC component described by \citet{poppe16} and \citet{nesvorny10} is spheroidal but centered at the Sun and its intensity decreases slowly with heliocentric distance at 1~AU.  While not isotropic when viewed from Earth orbit, the annual modulation of such a component is minimal.  We consider the OCC origin to be the most likely explanation of the amplitude measured by CIBER-NBS.

\section{Considerations for future measurements with this technique}

The achieved accuracy in the measurements presented here is largely limited by the short duration of a sounding rocket flight. Recently, a range of new opportunities for spaced-based small-aperture photometric measurements have arisen, most notably in the realm of cube-sats.  The NBS design is compact and could be done in the optical with un-cooled detectors.  

We recommend future experiments which seek to constrain the ZL intensity through Fraunhofer spectroscopy consider several lessons from our investigation.  If longer exposure times are available, pixels with smaller angular size should be implemented.  This would allow for deeper masking of stars, and reduced reliance on modeling of the stellar population, which can be a leading source of systematic error.  The resulting smaller instantaneous FOV would also help to reduce the contribution from ZL gradients. With more time in orbit, a greater number of fields at varied Galactic latitude can be targeted. Coadding results with varied DGL errors will allow for the reduction of uncertainty. 

Modern detector devices have demonstrated greatly reduced DC levels.  If low uncertainty can be obtained on residual DC, a narrow band measurement at 8520\AA~ could be deployed to constrain both the foregrounds and EBL in a single measurement.

%%%%%%%%%%%%%%%%%%%%%%%%%%%%%%%%%%%%%%%%%%%%%%%
\section{Acknowledgements}
The data presented here would not have been possible without the support of the NSROC staff at Wallops Flight Facility and White Sands Missile Range.  Insight from Kalevi Mattila and Roland dePutter improved the quality of the manuscript.  We thank Ned Wright for providing us with his model predictions and Leo Girardi for his star count code.  P.K. and M.Z. acknowledge support from the NASA postdoctoral program.  We'd also like to thank the late Keith Lykke, who's work along with Steven Brown and Allan Smith on the calibration program made this work possible. We acknowledge the contributions of collaboration members working early in the project that were important to acquiring the data in this paper including Ian Sullivan, Brian Keating and Tom Renbarger. This work was supported by NASA APRA research grants NNX07AI54G, NNG05WC18G, NNX07AG43G, NNX07AJ24G, NNX10AE12G, and NNX16AJ69G. Initial support was provided by an award to J.B. from the Jet Propulsion Laboratory’s Director’s Research and Development Fund. Japanese participation in CIBER was supported by KAKENHI (20·34, 18204018, 19540250, 21340047, and 21111004) from Japan Society for the Promotion of Science (JSPS) and the Ministry of Education, Culture, Sports, Science and Technology (MEXT). Korean participation in CIBER was supported by the Pioneer Project from Korea Astronomy and Space Science Institute (KASI).

\bibliography{NBS_results}
\bibliographystyle{apj}

\end{document}